\documentstyle[12pt]{article}
\begin{document}

\begin{center}
{\Large {\bf Gravitational instantons admit\\
       hyper-K\"{a}hler structure }}\\[10mm]
{\Large A. N. Aliev and Y. Nutku}\\[4mm] Feza G\"ursey Institute, P. K. 6
\c{C}engelk\"oy 81220 Istanbul, Turkey\\[10mm]
September 25, 1998
\end{center}

\noindent
{\large {\bf Abstract}}

  We construct the explicit form of three almost complex structures
that a Riemannian manifold with self-dual curvature admits and show
that their Nijenhuis tensors vanish so that they are integrable.
This proves that gravitational instantons with self-dual curvature
admit hyper-K\"{a}hler structure.
In order to arrive at the three vector valued $1$-forms defining almost
complex structure, we give a spinor description of real $4$-dimensional
Riemanian manifolds with Euclidean signature in terms of two independent sets
of $2$-component spinors. This is a version of the original Newman-Penrose
formalism that is appropriate to the discussion of the mathematical, as well
as physical properties of gravitational instantons. We shall build on the
work of Goldblatt who first developed an NP formalism for gravitational
instantons but we shall adopt it to differential forms in the NP basis to
make the formalism much more compact. We shall show that the
spin coefficients, connection $1$-form, curvature $2$-form, Ricci and
Bianchi identities, as well as the Maxwell equations naturally split
up into their self-dual and anti-self-dual parts corresponding to the two
independent spin frames. We shall give the complex dyad as well
as the spinor formulation of the almost complex structures and show that
they reappear under the guise of a triad basis for the Petrov classification
of gravitational instantons. Completing the work of Salamon on
hyper-K\"ahler structure, we show that the vanishing of the Nijenhuis
tensor for all three almost complex structures depends on the choice
of a self-dual gauge for the connection which is guaranteed by virtue of the
fact that the curvature $2$-form is self-dual for gravitational instantons.

\pagebreak

\section{Introduction}

    Gravitational instantons are described by $4$-dimensional real Riemannian
metrics with self-dual curvature. We shall prove that they admit rich complex
structure, namely hyper-K\"ahler structure. For a general metric with
self-dual curvature $2$-form we shall present the explicit
expressions for three vector valued $1$-forms that define almost complex
structure and further show that they are integrable as their Nijenhuis
tensors vanish. In physics literature \cite{gr} the
hyper-K\"ahler property of instantons was observed and used very
effectively in some important examples of gravitational instanton metrics,
whereas in the mathematical literature \cite{s} hyper-K\"ahler structure
is discussed in rather abstract terminology
of a category of ``words" which consist
of combinations of three letters generating elements of the space
of $2$-forms. Here we shall use a systematic approach which is based on the
Newman-Penrose formalism for Euclidean signature to provide
explicit proof of hyper-K\"ahler structure when curvature is self-dual.
Some of the results we shall present are not new, but they appear
as part of a systematic exposition in the framework of the
Newman-Penrose formalism which is a powerful technique that can be
applied to other problems of current interest in physics and mathematics.

  It is well-recognized that in general relativity the most useful formalism
for investigating properties of physically interesting exact solutions of
the Einstein field equations is the Newman-Penrose (NP) formalism \cite{np}
and this should be valid in the case of Euclidean signature as well.
The original NP formalism takes advantage of the
Lorentzian signature of spacetime in an essential way and is therefore
not suitable for studying properties of gravitational instantons \cite{gh},
\cite{egh} where the metric is strictly Riemannian. Goldblatt \cite{g1},
\cite{g2} has developed Newman-Penrose and Geroch-Held-Penrose formalisms
for Euclidean signature and we shall build on his work. The NP formalism
becomes compact when everything is expressed in terms of differential forms
appropriate to the NP basis, an approach
developed by one of us in an unpublished work \cite{old} some
time ago. Now we shall present the differential form version of the NP
formalism for Euclidean signature and in view of the fact that
it is largely even unknown for the Lorentzian case, we shall develope the
whole formalism {\it ab initio}. Finally, Goldblatt's work is
aimed towards the analysis of two-sided Ricci-flat metrics and he has
specialized the identities of Riemannian geometry to this end.
Our primary interest is in self-duality but we shall not specialize
the identities of the NP formalism.

The essential outlines of a Newman-Penrose formalism and the underlying
spinor structure of $4$-dimensional Riemannian manifolds were considered
by Penrose and Rindler \cite{pr} in their general consideration of complex
manifolds. Plebanski \cite{pl} has developed it for complex space-time.
These were followed up by the work of
Flaherty \cite{flaherty}, \cite{flaherty2} who discussed in detail
many of these issues, again, for $4$-dimensional complex manifolds.
Still another approach to complex spacetime is ${\cal H}$-space
\cite{newman}. We shall be interested in the reality condition that
selects real $4$-dimensional manifolds with Euclidean
signature and for this purpose it is more economical, in fact more
insightful, to bypass the detour through complex spaces altogether and
restrict the scope of our discussion from the outset to real manifolds.
This is also the approach of Goldblatt \cite{g1}. Earlier Gibbons and Pope
\cite{gp}, \cite{p} had used spinors in their discussion of gravitational
instantons without developing the full NP formalism.

Recently it was recognized that a class of gravitational instantons can
be put into one to one correspondence with minimal surfaces in 3-dimensional
Euclidean space \cite{n}, \cite{akn}.
In the case of Yang-Mills instantons such a correspondence was established
by Comtet \cite{comtet}. Among the gravitational instantons derived from
minimal surfaces, the instanton that corresponds to the helicoid minimal
surface has many interesting properties \cite{akhn} and the Newman-Penrose
formalism for positive definite signature is necessary in order to carry out
a full investigation of the properties of this as well as all gravitational
instantons.

The principal issue in the discussion of gravitational instantons is the
decomposition of the gravitational field into its self-dual and
anti-self-dual, alternatively left-half-flat and right-half-flat, parts and
this decomposition arises naturally if one appeals to spinors. Since the
Lorentz group is now replaced by $SO(4)$ which is isomorphic to
$ [\,SU(2)\times SU(2)\,]/Z_{2} $, {\it two independent} unitary
linear transformations of the complex $2$-dimensional plane
correspond to proper transformations of $4$-dimensional Euclidean space
and vice versa. Therefore we need two independent sets of spin frames for
a discussion of the spinor structure of $4$-dimensional Riemannian manifolds
with Euclidean signature. The necessity of dealing with two independent
spin frames may seem cumbersome but it has the added bonus that
in terms of these $2$-component spinors the self-dual and anti-self dual
properties of the gravitational field seperate out very naturally. We
shall show that the spin coefficients, connection $1$-form and
the curvature $2$-form split up into two {\it independent} sets
corresponding to the two spin frames and each set can be readily identified
as belonging to either one of the self-dual, or anti-self-dual sectors.

   Complex structure plays a vitally important role in real $4$-dimensional
Riemannian manifolds that describe gravitational instantons but this role is
often obscured by going through complex spacetime. We shall present the
explict expressions for three vector-valued $1$-forms that define almost
complex structures for $4$-dimensional real Riemannian manifolds and
furthermore show that they are all integrable when the curvature $2$-form
is self-dual. The condition for the integrability of these almost
complex structures, namely the vanishing of their Nijenhuis tensors, is
satisfied for a self-dual connection. In the case of gravitational
instantons this condition can always be satisfied by an appropriate choice
of gauge because curvature is self-dual. Hence we prove the theorem that
gravitational instantons admit tri-complex,
or hyper-K\"ahler structure. This property of gravitational instantons
was earlier used to great effect in particular examples \cite{gr}.
Goldblatt \cite{g1} has considered general expressions
for almost complex structure and proposed two vector-valued $1$-forms.
However, they cannot both define almost complex structure simultaneously
because one of them belongs to the self-dual, whereas the other
to the anti-self-dual sector and therefore only one of them
enters into hyper-K\"ahler structure. Salamon \cite{s} has discussed
hyper-K\"ahler structure, however, the proof of
integrability of the three almost complex structures through a check
of their Nijenhuis tensor is missing except in one trivial
case which follows directly from the K\"ahler property itself.
In sect.\ref{sec-complex} we shall present an explicit and systematic proof
of hyper-K\"ahler structure of gravitational instantons using the
Newman-Penrose formalism for Euclidean signature.

  In order to present a complete account we shall derive the Ricci and
Bianchi identities as well as the
Maxwell equations in terms of spin coefficients, complex dyad scalars
of curvature and electromagnetic field respectively, using the NP
basis differential forms. All these identities of Riemannian geometry fall
into two independent sets and each set can be obtained from the other by
an operation, ``tilde", that swaps the spin frames. So, in effect, with
the help of the tilde transformation we need to carry only one half of the
identities of Riemannian geometry. We shall also discuss the spinor
description of the topological invariants and Petrov classification
of gravitational instantons where the three self-dual basis $2$-forms
which are the K\"ahler $2$-forms for the three almost complex structures
in hyper-K\"ahler structure play the role of a triad basis. In the following
we shall use the work of Goldblatt \cite{g1} repeatedly, however, we shall
adopt it to differential forms which makes it compact and in the interest of
clarity we have found it useful to develope the whole formalism {\it ab
initio}.

\section{Complex dyad}

At each point on a 4-dimensional real Riemannian manifold we shall introduce
an isotropic dyad, a pair of complex $4$-vectors $l^\mu , m^\mu $ which
together with their complex conjugates serve to define a tetrad
\begin{equation}
\label{tetraddown}e_a^\mu =\{l^\mu \,,\bar l^\mu \,,m^\mu \,,\bar m^\mu \} 
\end{equation}
where bar denotes complex conjugation. The inverse of the basis
(\ref{tetraddown}) is $e_\nu ^a=\{\bar l_\nu
\,,l_\nu \,,\bar m_\nu \,,m_\nu \}$ so that the co-frame $1$-forms are given
by 
\begin{equation}
\label{1forms}\{\bar l\,,l\,,\bar m\,,m\}=e_{\;\nu }^adx^\nu =e^a 
\end{equation}
and the metric is expressed in the form 
\begin{equation}
\label{metric}ds^{\;2}=l\otimes \bar l+\bar l\otimes l+m\otimes \bar m+\bar
m\otimes m 
\end{equation}
which is the analog of the double null form in the Newman-Penrose formalism.
The metric has positive definite signature. The legs of the complex dyad satisfy
the normalization conditions
\begin{equation}
\begin{array}{c}
l_\mu \bar l^\mu =1,\;\;\;\;\;\;m_\mu \bar m^\mu =1, \label{norm1} \\
l_\mu l^\mu =0, \;\;m_\mu m^\mu =0, \;\;
 l_\mu m^\mu =0, \;\;l_\mu \bar{m}^\mu =0, \label{isotr}
\end{array}
\end{equation}
where eqs.(\ref{isotr}) express the fact that we have chosen an isotropic
dyad. In the isotropic frame the metric will be given by
\begin{equation}
\label{tetradcom}\eta _{ab}=\left(
\begin{array}{cccc}
0 & 1 & 0 & 0 \\
1 & 0 & 0 & 0 \\
0 & 0 & 0 & 1 \\
0 & 0 & 1 & 0
\end{array}
\right)
\end{equation}
which is manifest from (\ref{metric}).
Further, we have the completeness relation
\begin{equation}
\label{complete}\delta _{\;\nu }^\mu =l^\mu \bar l_\nu +\bar l^\mu l_\nu
+m^\mu \bar m_\nu +\bar m^\mu m_\nu 
\end{equation}
which follows from the choice of the complex dyad (\ref{tetraddown}) and its
inverse. Next we define the intrinsic derivative operators, or the
directional derivative along the legs of the complex dyad. Following
the Newman-Penrose notation we have
\begin{equation}
\label{operators}D=l^\mu \frac \partial {\partial x^\mu },\quad \qquad
\delta =m^\mu \frac \partial {\partial x^\mu } 
\end{equation}
together with their complex conjugates and inserting the completeness
relation (\ref{complete}) into the definition of the exterior derivative we
find 
\begin{equation}
\label{d}d=\bar l\,D+l\,\bar D+m\,\bar \delta +\bar m\,\delta 
\end{equation}
as the resolution of the exterior derivative along the legs of this complex dyad.

\section{Spin frames}

In the definition of the legs of the tetrad there is a degree of freedom, a
global gauge freedom, of rotation. This is given by $SO(4)$ which can be
decomposed into a product of two independent $SU(2)$ degrees of freedom. The
action of each $SU(2)$ can be described by a linear fractional
transformation of the compactified complex plane, that is, the Riemann
sphere and the spin frame is the familiar spinor basis. Thus we begin by
introducing two spin frames with bases 
\begin{equation}
\label{01}\zeta _{\,a}^A=\{\,o^A\,,\;\;\iota
^A\,\}\;\;\;\;\;\;\;A=1,2\;\;\;a=0,1 
\end{equation}
and 
\begin{equation}
\label{0'1'}{\tilde \zeta _{{x^{\prime }}}^{X^{\prime }}}=\{\,\tilde
o^{X^{\prime }}\,,\;\;\tilde \iota ^{X^{\prime }}\,\}\;\;\;\;\;\;\;X^{\prime
}=1^{\prime },2^{\prime }\;\;\;x^{\prime }=0^{\prime },1^{\prime } 
\end{equation}
of two independent $2$-component spinors respectively. Here capital Latin
indices refer to the components of spinors and small Latin indices run
over the two spinors, omicron and iota, respectively. Both sets of these
indices will be raised and lowered by the Levi-Civita symbol
from the right. Thus we have $o_A=o^B\epsilon _{BA}=-\epsilon _{AB}o^B$ and
the normalization conditions are 
\begin{equation}
\label{01contr}o_A\iota ^A=1,\;\;\;\;\;\tilde o_{X^{\prime }}\tilde \iota
^{X^{\prime }}=1 
\end{equation}
with all other contractions vanishing identically. It is not possible to
contract the unprimed and primed indices as they refer to objects that
belong to different spaces. It may seem redundant to use tilde over spinors
when this information is already carried by whether or not they carry primed
indices, however, as we shall soon see tilde is a very important operation
and it will be useful to keep it. Suitable combinations of spinors from the
bases (\ref{01}) and (\ref{0'1'}) will determine the complex dyad through the
Infeld-van der Waerden \cite{waerden} connecting quantities $\sigma
_{\;AX^{\prime }}^\mu $. Without loss of generality we may choose the
representation 
\begin{equation}
\label{delete}
\begin{array}{rcl}
l^\mu & = & \sigma _{\;00^{\prime }}^\mu =\sigma _{\;AX^{\prime }}^\mu
o^A\tilde o^{X^{\prime }} \\ 
\bar l^\mu & = & \sigma _{\;11^{\prime }}^\mu =\sigma _{\;AX^{\prime }}^\mu
\iota ^A\tilde \iota ^{X^{\prime }} \\ 
m^\mu & = & \sigma _{\;01^{\prime }}^\mu =\sigma _{\;AX^{\prime }}^\mu
o^A\tilde \iota ^{X^{\prime }} \\ 
\bar m^\mu & = & -\sigma _{\;10^{\prime }}^\mu =-\sigma _{\;AX^{\prime
}}^\mu \iota ^A\tilde o^{X^{\prime }} 
\end{array}
\end{equation}
which satisfies the normalization conditions (\ref{norm1}) and (\ref{01contr}%
). We may summarize these definitions in the form of a matrix 
\begin{equation}
\label{sigup1}\sigma _{\;AX^{\prime }}^\mu =\left( 
\begin{array}{cc}
l^\mu & m^\mu \\ 
-\bar m^\mu & \bar l^\mu 
\end{array}
\right) ,\qquad \qquad \sigma _\mu ^{\;AX^{\prime }}=\left( 
\begin{array}{cc}
\bar l_\mu & \bar m_\mu \\ 
-m_\mu & l_\mu 
\end{array}
\right) 
\end{equation}
which determine the correspondence between tensor and spinor fields. For
example, the components of the metric are given by 
\begin{equation}
\label{metcom}g_{\mu \nu }=\epsilon _{AB}\epsilon _{X^{\prime
}Y^{\prime }}\sigma _\mu ^{\;AX^{\prime }}\sigma _\nu ^{\;BY^{\prime }}
\end{equation}
and we shall write it simply as 
\begin{equation}
\label{short}
\begin{array}{rcl}
g_{\mu \nu } & \leftrightarrow & g_{AX^{\prime }BY^{\prime }}=\epsilon
_{AB}\epsilon _{X^{\prime }Y^{\prime }} \\  
& \Leftrightarrow & \epsilon _{AB} 
\end{array}
\end{equation}
following conventional usage. The Infeld-van der Waerden connecting
quantities satisfy 
\begin{equation}
\label{complete1}\delta _\mu ^{\;\nu }=\sigma _\mu ^{\;AX^{\prime }}\sigma
_{\;AX^{\prime }}^\nu ,\;\;\;\;\;\;\;\;\sigma _{\;AX^{\prime }}^\mu \sigma
_\mu ^{\;BY^{\prime }}=\delta _A^{\;B}\,\delta _{X^{\prime }}^{\;Y^{\prime
}} 
\end{equation}
which reproduce the completness relation (\ref{complete}). In dealing with $2
$-component spinors we shall repeatedly use the basic spinor identity in two
dimensions 
\begin{equation}
\label{basicid}\epsilon _{A\,[\,B}\;\epsilon _{C\,D\,]}=0 
\end{equation}
where square parantheses denote skew symmetrization. As a consequence we
have the familiar relation 
\begin{equation}
\label{basicid1}\xi _{[AB]}= \frac{1}{2}\,{\xi _C}^C \, \epsilon _{AB}
\end{equation}
where $\xi _{AB}$ is an arbitrary second rank 2-component spinor.

\section{Ricci rotation coefficients}

The Ricci rotation coefficients are the complex dyad components of the Levi-Civita
connection which are defined by 
\begin{equation}
\label{Rrc}\gamma _{ijk}=e_{j\,\mu ;\,\nu }\,e_{\;i}^\mu \,e_{\;k}^\nu
=-\gamma _{jik} 
\end{equation}
where semicolon denotes covariant differentiation. For the basis (\ref
{tetraddown}) the Ricci rotation coefficients will be labelled as 
\begin{equation}
\label{rotcoeff1}
\begin{array}{ll}
\kappa =\gamma _{311}=l_{\mu ;\nu }\,m^\mu l^\nu ,\qquad \qquad & \pi
=\gamma _{231}=m_{\mu ;\nu }\,\bar l^\mu l^\nu , \\ 
\tau =\gamma _{312}=l_{\mu ;\nu }\,m^\mu \bar l^\nu , & \nu =\gamma
_{232}=m_{\mu ;\nu }\,\bar l^\mu \bar l^\nu , \\ 
\sigma =\gamma _{313}=l_{\mu ;\nu }\,m^\mu m^\nu , & \mu =\gamma
_{233}=m_{\mu ;\nu }\,\bar l^\mu m^\nu , \\ 
\rho =-\gamma _{314}=l_{\mu ;\nu }\,m^\mu \bar m^\nu , & \lambda =-\gamma
_{234}=m_{\mu ;\nu }\,\bar l^\mu \bar m^\nu , 
\end{array}
\end{equation}
$$
\begin{array}{clcl}
\epsilon & =\frac 12(\gamma _{211}-\gamma _{341}) & = & \frac 12(l_{\mu ;\nu
}\,\bar l^\mu l^\nu -\bar m_{\mu ;\nu }\,m^\mu l^\nu ), \\[2mm]
\gamma & =\frac 12(\gamma _{212}+\gamma _{342}) & = & \frac 12(l_{\mu ;\nu
}\,\bar l^\mu \bar l^\nu +\bar m_{\mu ;\nu }\,m^\mu \bar l^\nu ), \\ [2mm]
\alpha & =-\frac 12(\gamma _{214}-\gamma _{344}) & = & -\frac 12(l_{\mu ;\nu
}\,\bar l^\mu \bar m^\nu -\bar m_{\mu ;\nu }\,m^\mu \bar m^\nu ), \\  [2mm]
\beta & =\frac 12(\gamma _{213}+\gamma _{343}) & = & \frac 12(l_{\mu ;\nu
}\,\bar l^\mu m^\nu +\bar m_{\mu ;\nu }\,m^\mu m^\nu ), 
\end{array}
$$
again following a notation as close as possible to the Newman-Penrose
spin coefficients for Lorentzian signature.

The most convenient way of calculating the Ricci rotation coefficients is
through the use of differential forms \cite{old}. Taking the exterior
derivative of the basis 1-forms (\ref{1forms}) and expressing the result in
terms of the basis 2-forms we obtain 
\begin{equation}
\label{dlm}
\begin{array}{lll}
d\,l & = & (\bar \gamma -\epsilon )\;l\wedge \bar l+(\alpha +\bar \beta
-\bar \pi )\;l\wedge m+(\tau -\beta -\bar \alpha )\;l\wedge \bar m \\  
&  & -\bar \nu \;\bar l\wedge m+\kappa \;\bar l\wedge \bar m-(\bar \lambda
+\rho )\;m\wedge \bar m \\[2mm]
d\,m & = & (\pi +\tau )\;\bar l\wedge l-(\bar \epsilon +\gamma -\lambda
)\;l\wedge m-\mu \;l\wedge \bar m \\  
&  & +(\epsilon -\rho +\bar \gamma )\;\bar l\wedge m+\sigma \;\bar l\wedge
\bar m-(\bar \alpha -\beta )\;m\wedge \bar m 
\end{array}
\end{equation}
where the coefficients are linear algebraic equations for the Ricci rotation
coefficients. Thus the exterior derivative of the basis 1-forms $l,m$ yields
the Ricci rotation coefficients by a simple comparison of the result with
eqs.(\ref{dlm}).

\section{Spin coefficients}

With the choice of two independent spin-frames we necessarily arrive at two
sets of spin coefficients which are defined by 
\begin{equation}
\label{spcoef}\Gamma _{ab\,AX^{\prime }}=\zeta _{a\,B;\,AX^{\prime }}\;\zeta
_b^B,\qquad \quad \tilde \Gamma _{x^{\prime }y^{\prime }\,AX^{\prime
}}=\tilde \zeta _{x^{\prime }\,Y^{\prime };\,AX^{\prime }}\;\tilde \zeta
_{y^{\prime }}^{\,Y^{\prime }} 
\end{equation}
where $ ;\mu \leftrightarrow \;;AX^{\prime }$ denotes the spinor
equivalent of the covariant derivative. This is a complex holomorphic
operator since the primed and unprimed indices are independent. We note the
symmetry in the first pair of complex dyad indices in both $\Gamma _{abAX^{\prime
}}$ and $\tilde \Gamma _{x^{\prime }y^{\prime }AX^{\prime }}$ so that their
traces will vanish. The spin coefficients are given in terms of the Ricci
rotation coefficients. It is clear that due to the symmetries of $\Gamma
_{abAX^{\prime }}$ and its tilde counterpart, one needs to evaluate only six
complex spin coefficients in each of the two cases. We find that $\Gamma
_{a\,\,\,cd^{\prime }}^{\;b}$ and $\tilde \Gamma _{x^{\prime
}\,\,\,cd^{\prime }}^{\;y^{\prime }}$, where indices are raised by the $2$%
-dimensional Levi-Civita symbol, can be listed according to the table:

\begin{center}
\begin{tabular}{|cc|c|c|c|c|c|c|c|}           \hline
      & $$ $b$  & $ $ $0$ & $ $ $1$ & $ $ $0$  
      & $$ $y'$ & $ $ $0'$& $ $ $1'$& $ $ $0'$ \\
      & $a$ $$  & $0$ $ $ & $0$ $ $ & $1$ $ $
      & $x'$ $ $& $0'$ $ $& $0'$ $ $& $1'$ $ $      \\
c$d'$ &         &         &         &
      &         &         &         &            \\       \hline
      &         &         &         &
      &         &         &         &             \\
0$0'$ &         & $\epsilon$ & $ -\kappa$ & $\bar{\tau}$
      &         &$-\bar{\gamma}$&$-\bar{\nu}$& $\pi$     \\
      &         &         &         &
      &         &         &         &            \\
1$0'$ &         & $ \alpha $  & $ -\rho  $ & $- \bar{\sigma} $
      &         &$\bar{\beta}$&$\bar{\mu}$ & $\lambda$  \\
      &         &         &         &
      &         &         &         &    \\
0$1'$ &         & $\bar{\alpha} $   & $ - \sigma $  & $ - \bar{\rho} $
      &         &$ \beta$           &$\bar{\lambda}$& $\mu$     \\
      &         &                   &            &
      &         &                   &            &   \\
1$1'$ &         & $ -\bar{\epsilon}$& $ - \tau $ & $ \bar{\kappa} $
      &         & $\gamma$          &$-\bar{\pi}$& $\nu$     \\
      &         &                   &            &
      &         &                   &            &   \\    \hline
\end{tabular}
\end{center}

\vspace{1mm}

\begin{center}
Table of Spin Coefficients
\end{center}

As we have remarked earlier, the most convenient way obtaining the spin
coefficients is through eqs.(\ref{dlm}).

\subsection{Swapping symmetry}

A glance at the table of spin coefficients suggests the introduction of
``swapping symmetry,'' namely tilde, which is an operation that turns
all quantities with tilde into those without and vice versa. In other words
tilde swaps the unprimed and primed spin frames,
alternatively the two $SU(2)$ components of $SO(4)$. The significance of the
tilde operation will emerge as the operation that turns self-dual into
anti-self-dual parts of the curvature $2$-form. Explicitly the tilde
operation is given by the replacement of either one of the complex dyad vectors $%
l\leftrightarrow \bar l$, or $m\leftrightarrow \bar m$. For the sake of
definiteness we shall use the tilde operation given by 
\begin{equation}
\label{tildesymmetry}l\leftrightarrow \bar l 
\end{equation}
with $m$ and $\bar m$ unchanged. In terms of the basic spinors this amounts
to the replacement 
\begin{equation}
\label{tildesymmetry1}o^A\mapsto i\,\tilde \iota ^{X^{\prime
}},\;\;\;\;\iota ^A\mapsto i\,\tilde o^{X^{\prime }},\;\;\;\;\tilde
o^{X^{\prime }}\mapsto -i\iota ^A,\;\;\;\;\tilde \iota ^{X^{\prime }}\mapsto
-io^A\;\;\;\; 
\end{equation}
which preserves the normalization conditions (\ref{01contr}). From eq.(\ref
{delete}) it is seen that that the tilde operation given in eqs.(\ref
{tildesymmetry}) - (\ref{tildesymmetry1}) in turn implies the correspondence 
\begin{equation}
\label{indexsymmetry}0\,\tilde \leftrightarrow \,1^{\prime }\;,
\;\;\;\;\;1\,\tilde
\leftrightarrow \,0^{\prime }
\end{equation}
between the spinor indices. The action of the tilde operation on spin
coefficients yields 
\begin{equation}
\label{tildesymmetrycoef}
\begin{array}{rclrclrcl}
\tau & \tilde \leftrightarrow & -\pi \;,\qquad \qquad & \epsilon & \tilde
\leftrightarrow & -\gamma \;,\qquad \qquad & \alpha & \tilde \leftrightarrow
& -\bar \beta \;, \\ 
\kappa & \tilde \leftrightarrow & -\nu \;, & \rho & \tilde \leftrightarrow & 
-\lambda , & \sigma & \tilde \leftrightarrow & -\mu 
\end{array}
\end{equation}
which is immediate from the definitions (\ref{rotcoeff1}). In fact, we could
have halved the names of spin coefficients by using the tilde symbol but it
is messy to deal with the complex conjugate of tilde. Finally, we note that
under the simultaneous interchange of the legs of the complex dyad $%
l\leftrightarrow \bar l,\;m\leftrightarrow \bar m$ the result is simply the
operation of complex conjugation which leaves the metric invariant.

\subsection{Commutator identities}

The commutator of different intrinsic derivative operators is given by the
Lie bracket 
\begin{equation}
\label{commut}\left[\,\partial_{j}, \partial_{k}\,\right]= {C^{\,i}}%
_{j\,k}\, \partial_{i}
\end{equation}
where the structure functions are given by 
$$
{C^{\,i}}_{j\,k} ={\gamma^{\,i}}_{k\,j} - {\gamma^{\,i}}_{j\,k}
$$
in terms of the Ricci rotation coefficients. Explicitly we have the
following set of commutator identities 
\begin{eqnarray}
D \bar{D} - \bar{D} D & = & (\bar{\epsilon}- \gamma)\, D +
(\bar{\gamma}- \epsilon)\, \bar{D} + (\bar{\pi} + \bar{\tau})\, \delta -
(\pi + \tau )\, \bar{\delta}
\label{com1} \\[2mm]
\delta \bar{\delta} - \bar{\delta} \delta & = & ( \lambda + \bar{\rho} )\, D
- ( \bar{\lambda}- \rho )\, \bar{D}  +   (\alpha - \bar{\beta})\, \delta -
(\bar{\alpha} - \beta )\, \bar{\delta}
\label{com2} \\[2mm]
D \delta - \delta  D & = & (\pi - \beta - \bar{\alpha} )\, D -
\kappa \,\bar{D} + (\epsilon + \bar{\gamma} - \bar{\lambda})\,\delta
- \sigma\, \bar{\delta}
\label{com3} \\[2mm]
\bar{D} \delta - \delta \bar{D} & = &  (\bar{\alpha} + \beta - \tau )\,
\bar{D} -  (\bar{\epsilon} + \gamma - \bar{\rho} )\,\delta + \nu \, D
+ \mu \, \bar{\delta}
\label{com4}
\end{eqnarray}
which along with their complex conjugates consist of all six identities as
the first two are pure imaginary. We note that under the tilde operation the
pure imaginary identities (\ref{com1}) and (\ref{com2}) remain invariant,
whereas (\ref{com3}) and (\ref{com4}) tranform into each other.

\section{Connection 1-forms}

The connection is an ${\cal SO}_4$-valued $1$-form which satisfies Cartan's
equations of structure 
\begin{equation}
\label{conform}de^\alpha +\omega _{\;\;\beta }^\alpha \wedge e^\beta =0, 
\end{equation}
where the matrix of connection $1$-forms is skew 
\begin{equation}
\label{skewom}\omega _{\alpha \beta }=-\ \omega _{\beta \alpha } 
\end{equation}
and due to this anti-symmetry, using the basic identity (\ref{basicid1}),
the spinor equivalent of the full connection $1$-form 
\begin{equation}
\label{conspform}
\begin{array}{rcl}
\omega _{\alpha \beta } & \leftrightarrow & \Gamma _{ax^{\prime
}\;\;by^{\prime }} \\  
&  & =\Gamma _{ab}\,\epsilon _{x^{\prime }y^{\prime }}+\tilde \Gamma
_{x^{\prime }y^{\prime }}\,\epsilon _{ab} \\  
& \Leftrightarrow & \Gamma _{ab} 
\end{array}
\end{equation}
naturally splits up into two pieces appropriate to the two sets of spin
frames. Here $\Gamma _{ab}$ and $\tilde \Gamma _{x^{\prime }y^{\prime }}$
are symmetric matrices of $1$-forms and raising an index with the
appropriate $2$-dimensional Levi-Civita symbol we have 
\begin{equation}
\label{pot}\Gamma _a^{\;\;b}=\Gamma _{a\;\;cx^{\prime }}^{\;\;b}\;\sigma
_\mu ^{cx^{\prime }}\,dx^\mu ,\qquad \quad \quad \tilde \Gamma _{x^{\prime
}}^{\;\;y^{\prime }}= \tilde \Gamma _{x^{\prime }\;\;az^{\prime }}^{\;\;y^{\prime
}}\;\sigma _\mu ^{az^{\prime }}dx^\mu 
\end{equation}
which are traceless. From eqs.(\ref{spcoef}), (\ref{sigup1}), (\ref
{rotcoeff1}) and the table of spin coefficients we find that 
\begin{equation}
\label{potform}
\begin{array}{l}
\Gamma _0^{\;\;0}=\frac 12\left( l_{\mu ;\nu }\ \bar l^\mu +m_{\mu ;\nu }\
\bar m^\mu \right) dx^\nu =\epsilon \,\bar l-\bar \epsilon \,l-\alpha
\,m+\bar \alpha \,\bar m \\ 
[2mm] \Gamma _0^{\;\;1}=-l_{\mu ;\nu }\ m^\mu dx^\nu =-\tau \,l-\kappa
\,\bar l+\rho \,m-\sigma \,\bar m \\ 
[2mm] {\Gamma }_1^{\;\;0}=-\bar \Gamma _0^{\;\;1},\quad \qquad \qquad \Gamma
_1^{\;\;1}=-\Gamma _0^{\;\;0} 
\end{array}
\end{equation}
and 
\begin{equation}
\label{potform'}
\begin{array}{l}
\tilde \Gamma _{0^{\prime }}^{\;\;0^{\prime }}=\frac 12\left( l_{\mu ;\nu
}\bar l^\mu -m_{\mu ;\nu }\bar m^\mu \right) dx^\nu =\gamma \,l-\bar \gamma
\,\bar l+\beta \,\bar m-\bar \beta \,m \\ 
[2mm] \tilde \Gamma _{0^{\prime }}^{\;\;1^{\prime }}=l_{\mu ;\nu }\ \bar
m^\mu dx^\nu =-\bar \pi \,l-\bar \nu \,\bar l-\bar \mu \,m+\bar \lambda
\,\bar m \\ 
[2mm] \tilde \Gamma _{1^{\prime }}^{\;\;0^{\prime }}=-\overline{\widetilde{%
\Gamma }}\ _{0^{\prime }}^{\;\;1^{\prime }},\qquad \quad \qquad \widetilde{%
\Gamma }_{1^{\prime }}^{\;\;1^{\prime }}=-\widetilde{\Gamma }_{0^{\prime
}}^{\;\;0^{\prime }} 
\end{array}
\end{equation}
provide the explicit expression for the connection $1$-forms. There are
important differences here with the Newman-Penrose formalism for Lorentzian
signature. First of all we have two sets of connection $1$-forms
in place of a single set. These two sets of connection $1$-forms
necessarily exhibit a ``redundancy'' in that their off-diagonal
elements are related. This is expected since we are dealing with two
spin frames and setting up two sets of connection $1$-forms, whereas
the number of independent spin coefficients is fixed by the number of
dimensions of the manifold. This redundancy is actually a blessing
in disguise because eventually it will prove to be an
enormous convenience in our considerations of self-dual
and anti-self-dual $2$-forms. Finally, we observe that
\begin{equation}
\label{dsig}d\sigma ^{ax^{\prime }}=-\Gamma _b^{\;\;a}\wedge \sigma
^{bx^{\prime }}-\tilde \Gamma _{y^{\prime }}^{\;\;x^{\prime }}\wedge \sigma
^{ay^{\prime }} 
\end{equation}
is the compact form of eqs.(\ref{dlm}). Under the tilde operation $\Gamma $
simply goes over into $\tilde \Gamma $ according to the rules (\ref
{tildesymmetry}) - (\ref{tildesymmetrycoef}).

\subsection{Self-dual gauge}

A most convenient but not necessary device in dealing with $4$-dimensional
Riemannian manifolds that admit (anti)-self-dual curvature is to pick a
frame where the connection itself shares this property. We shall now show
that the two independent sets of connection $1$-forms in eqs.(\ref{potform})
and (\ref{potform'}) determine the self-dual and anti-self-dual connection
$1$-forms. With the definition
\begin{equation}
\label{sdconn5}^{\pm }\omega _{\alpha \beta }=\frac 12\left( \omega _{\alpha
\beta }\pm \frac{1}{2}{\epsilon _{\alpha \beta }}^{\gamma \delta }\,\omega
_{\gamma \delta }\right) , 
\end{equation}
the self-dual and anti-self dual sets of connection $1$-forms are given by 
\begin{equation}
\label{sdconn6}^{-}\omega _{\alpha \beta }=0,\qquad \qquad
^{+}\omega_{\alpha \beta }=0, 
\end{equation}
respectively. For the spinor equivalent of $^{\pm }\omega _{\alpha \beta }$
we need the spinor equivalent of the totally anti-symmetric Levi-Civita
alternating symbol 
\begin{equation}
\label{lcspin'}\epsilon _{\alpha \beta \gamma \delta }\leftrightarrow
\epsilon _{ax^{\prime }by^{\prime }cz^{\prime }dw^{\prime }}=\epsilon
_{ac}\,\epsilon _{bd}\,\epsilon _{x^{\prime }w^{\prime }}\,\epsilon
_{y^{\prime }z^{\prime }}\,-\,\epsilon _{ad}\,\epsilon _{bc}\,\epsilon
_{a^{\prime }z^{\prime }}\,\epsilon _{b^{\prime }d^{\prime }} 
\end{equation}
which enters into eq.(\ref{sdconn5}) and we find 
\begin{equation}
\label{dualconsp}
\begin{array}{c}
^{-}\omega _{ax^{\prime }by^{\prime }}= \; ^{-}\omega
_{(ab)[x^{\prime}y^{\prime }]}= \Gamma _{ab}\epsilon _{x^{\prime }y^{\prime
}}, \\
^{+}\omega _{ax^{\prime }by^{\prime }}=\; ^{+}\omega
_{[ab](x^{\prime}y^{\prime })}= \Gamma _{x^{\prime }y^{\prime }}\epsilon
_{ab} 
\end{array}
\end{equation}
respectively. We conclude that the necessary and sufficient conditions for
the connection $1$-form to be self-dual, or anti-self-dual are given by 
\begin{equation}
\label{sdcond}\Gamma _{ab}\equiv 0,\qquad \qquad \tilde \Gamma _{x^{\prime
}y^{\prime }}\equiv 0 
\end{equation}
respectively. Returning back to eqs.(\ref{potform}) and (\ref{potform'}) we
note that the condition for self-duality of the connection implies 
\begin{equation}
\label{sdspcoef}\epsilon =\alpha =\tau =\kappa =\rho =\sigma =0, 
\end{equation}
while the criterion for anti-self-duality requires that 
\begin{equation}
\label{asdspcoef}\gamma =\beta =\pi =\mu =\nu =\lambda =0. 
\end{equation}
Given a metric with (anti)-self-dual curvature $2$-form it is always
possible to choose a gauge, namely frame for which either eqs.(\ref{sdspcoef}%
), or (\ref{asdspcoef}) hold.

\section{Basis 2-forms}

The spinor equivalent of the basis $2$-forms are wedge products of the
Infeld-van der Waerden matrices of basis $1$-forms (\ref{sigup1}) 
$$
e^a_{\mu}\, e^b_{\nu} \,dx^\mu \wedge dx^\nu \;\;\leftrightarrow
\;\;\sigma^{ax^{\prime }} \wedge \sigma ^{by^{\prime }} 
$$
and we have two sets of basis 2-forms 
\begin{eqnarray}
L_a^{\;\;b} &  = & \frac{1}{2} \,\sigma_{\mu \,ax^{\prime }}\,
 \sigma _\nu ^{\;\;bx^{\prime }}\,dx^\mu \wedge
dx^\nu  =  \frac{1}{2}\, \sigma _{ax^{\prime }}\wedge \sigma ^{bx^{\prime }}
\label{deflab00} \\
\tilde L_{x^{\prime }}^{\;\;y^{\prime }} & = & \frac{1}{2}\,
\sigma _{ax^{\prime }}\wedge
\sigma ^{ay^{\prime }} \label{deflab00000}
\end{eqnarray}
corresonding to the two spin frames. These are consequences of the spinor
relation 
\begin{equation}
\label{bfexp}L_{ax^{\prime}}^{\;\;by^{\prime}} = \sigma _{ax^{\prime
}}\wedge \sigma ^{by^{\prime }} = L_{a}^{\;\;b} \, \epsilon _{x^{\prime
}}^{\;\;y^{\prime }} + \tilde L_{x^{\prime }}^{ \;\;y^{\prime }}\,
\epsilon_a^{\;\;b} 
\end{equation}
due to skew symmetry in $ax^{\prime} , \; by^{\prime} $ and the basic spinor
identity (\ref{basicid1}). Taking the components of eqs.(\ref{deflab00}) and
(\ref{deflab00000}) we find
\begin{equation}
\label{sdbasis}
\begin{array}{rclrcl}
L_0^{\;\;0} & = & \frac 12\,\left( l\wedge \bar l+m\wedge \bar m\right)
,\qquad & \tilde L_{0^{\prime }}^{\;\;0^{\prime }} & = & \frac 12\,\left(
l\wedge \bar l-m\wedge \bar m\right) , \\ 
L_0^{\;\;1} & = & -l\wedge m, & \tilde L_{0^{\prime }}^{\;\;1^{\prime }} & =
& l\wedge \bar m, \\ 
L_1^{\;\;0} & = & \bar l\wedge \bar m, & \tilde L_{1^{\prime
}}^{\;\;0^{\prime }} & = & -\bar l\wedge m 
\end{array}
\end{equation}
and the remaining components follow from the vanishing of the trace. We note
that once again ${L_0}^1=-{\bar L_1}^0$ and $\tilde L_{0^{\prime
}}^{\;\;1^{\prime }} = - \bar{\tilde L}_{1^{\prime}}^{\;\;0^{\prime }} $ as
in the case of eqs.(\ref{potform}), (\ref{potform'}) and this property will
hold for the curvature $2$-forms as well.

The decomposion of the basis $2$-forms into self-dual and anti-self-dual
parts is immediate. Using the definition of Hodge star and
the completeness relation (\ref{complete}), we find that
\begin{equation}
\label{hodgelab}^{*}L_a^{\;\;b}= - L_a^{\;\;b}\;\;\;\;\;\;\qquad ^{*}\tilde
L_{x^{\prime }}^{\;\;y^{\prime }}=  \tilde L_{x^{\prime }}^{\;\;y^{\prime }}
\end{equation}
so that the two sets of independent basis $2$-forms (\ref{deflab00}), (\ref
{deflab00000}) can be recognized as the self-dual and anti-self-dual
objects. Finally, we note that these $2$-forms satisfy 
\begin{equation}
\label{covder1}
\begin{array}{rcl}
d\ L_a^{\;\;b}\;+\;L_a^{\;\;c}\wedge \Gamma _c^{\;\;b}\;-\;\Gamma
_a^{\;\;c}\wedge L_c^{\;\;b} & = & 0 \\ 
[2mm] d\ \tilde{L}{_{x^{\prime }}}^{\;y^{\prime }}+ \tilde{L}{_{x^{\prime }}}%
^{\;\;z^{\prime }}\wedge {\tilde \Gamma _{z^{\prime }}}^{\;\;y^{\prime }}-{%
\tilde \Gamma _{x^{\prime }}}^{\;\;z^{\prime }}\wedge \tilde{L}_{z^{\prime
}}^{\;\;y^{\prime }} & = & 0 
\end{array}
\end{equation}
which may be called the ``zeroth" Bianchi identities but are better known as
Ricci's lemma. Their verification is immediate from eqs.(\ref{sdbasis}) and (%
\ref{potform})-(\ref{potform'}). Hence $L_a^{\;\;b}$ and $\tilde
L_{x^{\prime }}^{\;\;y^{\prime }}$ form the spinor equivalent of the
self-dual and anti-self-dual basis of the space of $2$-forms respectively
and their covariant derivatives vanish identically. In eqs.(\ref{covder1})
we see a phenomenon which will appear repeatedly, namely, we need to write
down only one half of the ``zeroth'' Bianchi identities and remember that
the same equation is valid for its tilde version with primed indices as
well. In other words these equations hold for both sets of spin frames
independently.

\section{Complex structure}
\label{sec-complex}

   The complex dyad formalism and the corresponding spinor structure for
$4$-dimension\-al real strictly Riemannian manifolds provides a very natural
framework for the discussion of the complex structure of gravitational
instantons in its full generality. A real even-dimensional manifold will
admit almost complex structure provided there exists a real differentiable
vector-valued $1$-form
\begin{equation}
\label{almostcomplex}J=J_{\mu} ^{\;\;\nu} \;dx^\mu \otimes
\frac \partial {\partial x^\nu }
\end{equation}
which is an anti-involution with the components of $J$ satisfying 
\begin{equation}
\label{squarem1}J_\mu ^{\;\;\;\rho }J_\rho ^{\;\;\;\nu }=-\delta _\mu ^\nu
\end{equation}
so that 
\begin{equation}
\label{squarem12}J[J(\xi )]=-\xi 
\end{equation}
for any differentiable vector field $\xi $. The structure functions of
the almost complex structure must be real, that is
\begin{equation}
\label{hermite}J_\mu ^{\;\;\alpha }\;J_\nu ^{\;\;\beta }\ \ g_{\alpha \beta
}=g_{\mu \nu } 
\end{equation}
which is the Hermitian property. From eqs.(\ref{squarem1}) and
(\ref{hermite}) it follows that lowering the vector index with the
metric we get a skew-symmetric tensor
\begin{equation}
\label{kahler2comp}J_{\mu \nu }=-J_{\nu \mu } 
\end{equation}
and in this connection it is worth recalling that almost complex structure
is a metric-independent concept but of course in our case there exists a
Riemannian metric and we shall use it.
                            
   The explicit expression for almost complex structure of a general
gravitational instanton metric that admits self-dual curvature $2$-form
assumes a simple form in the complex dyad (\ref{tetraddown}). There exists
three such vector valued $1$-forms, which is implicit in Salamon \cite{s},
that together will form hyper-K\"ahler structure, namely
\begin{eqnarray}
J_{1\;\mu }^{\;\;\;\;\;\nu}& = & - i \, \left(l_\mu \, m^\nu - m_\mu l^\nu
- \bar{l}_\mu \, \bar{m}^\nu + \bar{m}_\mu \bar{l}^\nu \right)
\nonumber \\[1.5mm]
J_{2\;\mu }^{\;\;\;\;\;\nu}& = & l_\mu \, m^\nu - m_\mu l^\nu
+ \bar{l}_\mu \, \bar{m}^\nu - \bar{m}_\mu \bar{l}^\nu
\label{complexbizim}\\ [1.5mm]
J_{3\;\mu }^{\;\;\;\;\;\nu}& = & -i\left( l_\mu \bar l^\nu - \bar
l_\mu l^\nu + m_\mu \bar m^\nu - \bar m_\mu m^\nu \right) \nonumber
\end{eqnarray}
which can readily be verified to be real and satisfy all the properties
required of an almost complex structure, as expressed by eqs.(\ref{squarem1}),
(\ref{hermite}) and (\ref{kahler2comp}). Furthermore, they transform
properly under $SU(2)$ whereas in general only tensorial transformation
properties under $SO(4)$ would be required. Hyper-K\"ahler structure where
\begin{equation}
J_{i\;\mu }^{\;\;\;\;\;\sigma} J_{j\;\sigma}^{\;\;\;\;\;\nu} = -
J_{j\;\mu }^{\;\;\;\;\;\sigma} J_{i\;\sigma}^{\;\;\;\;\;\nu} =
\epsilon_{ijk} \,  J_{k\;\mu }^{\;\;\;\;\;\nu}
\label{hyp}
\end{equation}
is responsible for this. Goldblatt \cite{g1} had proposed the almost complex
stucture $J_{3\;\mu }^{\;\;\;\;\;\nu}$ above and another one that differs
from it by the choice of opposite sign in the last two terms. However,
these two choices belong to different sectors of self-duality and they
cannot coexist in hyper-K\"ahler structure.

The structure functions of
these three almost complex structures can be used to define $2$-forms
$ \omega_i = \frac{1}{2} J_{i\;\;\mu \nu }\,dx^\mu \wedge dx^\nu $, namely
\begin{eqnarray}
\omega_1 & = & - i \, (l\wedge m - \bar l \wedge \bar m ) \nonumber\\[1.5mm]
\omega_2 & = & l\wedge m + \bar l \wedge \bar m  \label{kahler2f}\\  [1.5mm]
\omega_3 & = & - i \,(l\wedge \bar l+m\wedge \bar m) \nonumber
\end{eqnarray}
which are simply the three real self-dual basis $2$-forms of
eqs.(\ref{sdbasis}). In particular, $\omega_3$ is the K\"ahler $2$-form.
We have already remarked that the connection
$1$-form for gravitational instantons can be chosen such that the
gauge is self-dual which is given by eqs.(\ref{sdspcoef}). Thus
the $2$-forms (\ref{kahler2f}) are closed by virtue of ``zeroth"
Bianchi identities (\ref{covder1}), or this can be demonstrated explicitly
using eqs.(\ref{dlm}). So from the structure functions of the three almost
complex structures we obtain the hyper-K\"ahler $2$-forms.
The spinor equivalent of almost complex structure assumes the form
\begin{equation}
J_{i \; \mu \nu }  \leftrightarrow
J_{i \; AB }\,\epsilon_{X^{\prime }Y^{\prime }} +
\tilde J_{i \; X'Y' } \,\epsilon_{AB}
\label{speqalcom}
\end{equation}
where the essential element, namely the second rank symmetric spinors
$J_{i \; AB }$ and $\tilde J_{i \; X'Y' } $ are simple bi-spinors
\begin{eqnarray}
J_{1\;AB}& = & -i \,(\, o_A o_B  -\iota_A \iota_B \,)
 \nonumber \\[2mm]
J_{2 \;AB} & = &  o_A o_B  + \iota_A \iota_B
\\  [2mm]
J_{3 \;AB} & = & 2 \,i \, o_{(A} \iota_{B)}   \nonumber
\label{cssp}
\end{eqnarray}
constructed from the basis spinors. Similar relations hold for primed
indices.

The question now arises as to whether or not these almost complex structure
are integrable which is the necessary and sufficient condition for
hyper-K\"ahler structure. The integrability condition is given by
the vanishing of the Nijenhuis tensor \cite{nijen} 
\begin{equation}
\label{nijenhuis}
N_{\mu \nu }^{\;\;\alpha} =   J_{\mu}^{\;\;\sigma}
(J_{\nu ;\sigma }^{\;\;\;\alpha} -J_{\sigma ;\nu }^{\;\;\;\alpha} )
- J_{\nu}^{\;\;\sigma} (J_{\mu ;\sigma }^{\;\;\;\alpha} -
J_{\sigma ;\mu }^{\;\;\;\alpha} )
\end{equation}
which is a vector-valued $2$-form. It will be convenient to define both a
vector-valued $1$-form ${\cal J}^\mu $ and a form-valued vector field ${\bf J%
}_\nu $ 
\begin{equation}
\label{nijencomp}{\cal J}^\mu \equiv J_{\;\nu }^\mu \,dx^\nu ,\qquad \quad 
{\bf J}_\nu \equiv J_{\;\nu }^\mu \,\frac \partial {\partial x^\mu } 
\end{equation}
in terms of which the Nijenhuis vector-valued $2$-form can be written as 
\begin{equation}
\label{nijen2}N^\alpha = \frac{1}{2} N_{\;\mu \nu }^\alpha \, dx^\mu \wedge
dx^\nu =dx^\nu \wedge {\it i}_{{\bf J}_\nu }\left\{ d\,{\cal J}^\alpha
\right\} 
\end{equation}
where ${\it i}$ denotes contraction of the $2$-form $d{\cal J}^\alpha $ with
the vector field ${\bf J}_\nu $. The Nijenhuis tensor is metric-independent
because the connection washes out in eq.(\ref{nijenhuis}) so that all the
covariant derivatives can be replaced by partial derivatives.

In our case the three Nijenhuis vector-valued $2$-forms are given by
\begin{eqnarray}
 N_{1}^{\;\alpha} & = &
\left[\, ( \bar{\rho} + 2\,\bar{\epsilon} -\sigma) \, l^\alpha
+( \kappa +2\,\alpha +\bar{\tau} ) \, m^{\alpha} \,\right] \,
( l\wedge \bar{l} +  m\wedge \bar{m} )
\nonumber \\  & &
+ \left[\, ( \tau + \bar{\kappa} +2\,\bar{\alpha})\, l^\alpha + (\bar{\sigma} -\rho -
2\,\epsilon) \, m^\alpha \, \right]\,( l\wedge m + \bar{l} \wedge \bar{m})
 + cc \nonumber \\ [2mm]
N_{2}^{\;\alpha} & = & \left[\, ( \bar{\rho} + 2\,\bar{\epsilon} + \sigma)
\, l^\alpha +(2\,\alpha +\bar{\tau}- \kappa ) \, m^{\alpha} )\,\right] \,
( l\wedge \bar{l} +  m\wedge \bar{m} ) \label{nexplicit} \\ & &
+ \left[\, ( \bar{\kappa} - \tau  - 2\,\bar{\alpha})\,l^\alpha
+ (\rho + \bar{\sigma}  + 2\,\epsilon) \, m^\alpha \,\right]\,
( l\wedge m  - \bar{l} \wedge \bar{m})
 + cc \nonumber   \\ [2mm]
 N_{3}^{\;\alpha} & = &
4\,( \bar{\kappa} \, l^\alpha + \bar{\sigma} \, m^{\alpha} ) \,l\wedge m
+ cc   \nonumber
\end{eqnarray}
where $ cc $ denotes the complex conjugate of the foregoing terms
and the condition for all of these almost complex structures to
be integrable requires that the coefficients of vector valued $2$-forms
in eqs.(\ref{nexplicit}) must all vanish. As it was noted by Salamon
\cite{s}, $N_{3}^{\;\alpha} = 0$ is automatically satisfied for K\"ahler
metrics, however, he has not directly checked the vanishing of
the Nijenhuis tensor for the remaining two almost complex structures.
From eqs.(\ref{nexplicit}) we arrive the conditions 
\begin{equation}
\label{nijensart} \epsilon =\alpha =\tau =\kappa =\rho =\sigma =0,
\end{equation}
for hyper-K\"ahler structure. But this is precisely the same as
the condition for a self-dual connection given by eqs.(\ref{sdspcoef}).
Since gravitational instantons admit self-dual curvature
we can always find a frame where the connection is self-dual. Thus
we have

\vspace{1mm}

\noindent
{\bf Theorem}

{\it Gravitational instantons admit hyper-K\"ahler, or tri-complex
structure.} The almost complex structures given by eqs.(\ref{complexbizim})
are all integrable for gravitational instantons.

\section{Curvature}

 Complex dyad scalars formed from the components of the Riemann curvature
tensor in the frame (\ref{tetraddown}) are physically important quantities.
As in the Newman-Penrose formalism with Lorentzian signature it will be
useful to start with a list of curvature scalars adopted to this tetrad.
First we have the usual decomposition of the Riemannian tensor
\begin{eqnarray}
R_{\mu\nu\rho\sigma} & = & C_{\mu\nu\rho\sigma}
+ \frac{1}{2} \,\left( g_{\mu\rho} R_{\nu\sigma} -
 g_{\mu\sigma} R_{\nu\rho} + g_{\nu\sigma} R_{\mu\rho}
 -  g_{\nu\rho} R_{\mu\sigma} \right) \nonumber \\ & &
 -  \frac{1}{6} \, ( g_{\mu\rho} g_{\nu\sigma}-  g_{\mu\sigma} g_{\nu\rho}) R
\label{riemdecom}
\end{eqnarray}
where $C_{\mu \nu \rho \sigma }$ is the conformal Weyl tensor, $R_{\mu \nu
}=g^{\rho \sigma }R_{\mu \rho \nu \sigma }$ is the Ricci tensor and $R$ is
the scalar of curvature. For the complex dyad (\ref{tetraddown}) we introduce two
sets of Weyl scalars 
\begin{equation}
\label{weylsc1}
\begin{array}{ll}
\Psi _0=C_{1313}=C_{\mu \nu \sigma \tau }l^\mu m^\nu l^\sigma m^\tau \qquad
& \tilde \Psi _0=C_{1414}=C_{\mu \nu \sigma \tau } l^\mu \bar m^\nu l^\sigma
\bar m^\tau \\ 
[2mm] \Psi _1=C_{1213}=C_{\mu \nu \sigma \tau }l^\mu \bar l^\nu l^\sigma
m^\tau & \tilde \Psi _1=C_{1241}=C_{\mu \nu \sigma \tau }l^\mu \bar l^\nu
\bar m^\tau l^\sigma \\ 
[2mm] \Psi _2=C_{1324}=C_{\mu \nu \sigma \tau }l^\mu m^\nu \bar l^\sigma
\bar m^\tau & \tilde \Psi _2=C_{1423}=C_{\mu \nu \sigma \tau }l^\mu \bar
m^\nu \bar l^\sigma m^\tau 
\end{array}
\end{equation}
which consist of a total of ten complex dyad scalars as Weyl scalars with
subscript $2$ are real. In addition we also introduce the trace-free Ricci
scalars 
\begin{equation}
\label{npnetrod}
\begin{array}{llrl}
\Phi _{00} & =-\frac 12\,R_{\mu \nu }l^\mu l^\nu =\bar \Phi _{22},\qquad & 
\Phi _{01} & =-\frac 12\,R_{\mu \nu }l^\mu m^\nu =-\bar \Phi _{21}, \\ 
[2mm] \Phi _{02} & =-\frac 12\,R_{\mu \nu }m^\mu m^\nu =\bar \Phi _{20}, & 
\Phi _{10} & =\frac 12\,R_{\mu \nu }l^\mu \bar m^\nu =-\bar \Phi _{12}, \\ 
[2mm] \Phi _{11} & =-\frac 12\,R_{\mu \nu }l^\mu \bar l^\nu +3\Lambda &  &  
\end{array}
\end{equation}
where $\Phi _{11}$ is real and thus we have the nine independent trace-free
Ricci scalars. Together with the scalar of curvature 
\begin{equation}
\label{scalarcurv}\Lambda =\frac 1{24}R 
\end{equation}
the complex dyad scalars (\ref{weylsc1}) and (\ref{npnetrod}) determine the
required twenty independent components of the Riemann tensor.

Following the general scheme of Penrose we shall obtain the curvature
spinors which are the spinor equivalent of the Riemann curvature tensor (\ref
{riemdecom}). The decomposition of the Riemann curvature spinor into Weyl,
trace-free Ricci and the scalar of curvature spinors in terms of 2-component
spinors brings out the meaning of the curvature complex dyad scalars in the
clearest way. Using the symmetry properties of the Riemann tensor, repeated
applications of the basic spinor identity (\ref{basicid1}) yield 
\begin{equation}
\label{curdecomsp}
\begin{array}{lll}
R_{\alpha \beta \gamma \delta } & \leftrightarrow & R_{AX^{\prime
}BY^{\prime }CZ^{\prime }DW^{\prime }} \\  
&  & =X_{ABCD}\;\epsilon _{X^{\prime }Y^{\prime }}\;\epsilon _{Z^{\prime
}W^{\prime }}+\tilde X_{X^{\prime }Y^{\prime }Z^{\prime }W^{\prime
}}\;\epsilon _{AB}\;\epsilon _{CD} \\  
&  & \hspace{5mm}+ \Phi _{ABZ^{\prime }W^{\prime }}\;\epsilon _{X^{\prime
}Y^{\prime }}\;\epsilon _{CD}+\tilde \Phi _{X^{\prime }Y^{\prime
}CD}\;\epsilon _{AB}\;\epsilon _{Z^{\prime }W^{\prime }} 
\end{array}
\end{equation}
where we have introduced the curvature spinors 
\begin{equation}
\label{cursp1}X_{ABCD}=\frac 14\,R_{ABM^{\prime }\;\;CDN^{\prime
}}^{\;\;\;\;\;\;\;\;\;\;M^{\prime }\;\;\;\;\;\;\;N^{\prime
}},\;\;\;\;\;\;\;\;\;\Phi _{ABX^{\prime }Y^{\prime }}=\frac
14\,R_{ABM^{\prime }\;\;N\;\;\;X^{\prime }Y^{\prime
}}^{\;\;\;\;\;\;\;\;\;\;M^{\prime }\;\;N} 
\end{equation}
together with their tilde counterparts. The anti-symmetry of the curvature
tensor in its first and last pair of indices gives rise to the relations 
\begin{equation}
\label{simsp1}X_{ABCD}=X_{(AB)(CD)},\qquad \Phi _{ABX^{\prime }Y^{\prime
}}=\Phi _{(AB)(X^{\prime }Y^{\prime })} 
\end{equation}
and similarly for the tildes. The symmetry of the Riemann tensor under the
interchange of the first and last pair of indices is reflected by 
\begin{equation}
\label{simsp2}X_{ABCD}=X_{CDAB},\qquad \Phi _{ABX^{\prime }Y^{\prime
}}=\tilde \Phi _{ABX^{\prime}Y^{\prime }} 
\end{equation}
in the the curvature spinors. It follows that only the curvature spinors $%
X_{ABCD}$, $\tilde X_{X^{\prime}Y^{\prime }Z^{\prime }W^{\prime }}$ and $%
\Phi _{ABX^{\prime }Y^{\prime }}$ are independent. As a consequenses of
these relations we have 
\begin{equation}
\label{conrel}{X_{ABC}}^B=3\Lambda \,\epsilon _{AC} 
\end{equation}
and $\Lambda =\tilde \Lambda$, {\it cf} eq.(\ref{scalarcurv}). Using these
results and also the basic spinor identity (\ref{basicid1}) we can write the
totally symmetric part of $X_{ABCD}$ as follows 
\begin{equation}
\label{weylsp}X_{ABCD}=\Psi _{ABCD}+\Lambda \,(\epsilon _{AC}\,\epsilon
_{BD}+\epsilon _{AD}\,\epsilon _{BC}),\qquad \Psi _{ABCD}=\Psi _{(ABCD)}
\label{totsym}
\end{equation}
and similarly for $\tilde X_{X^{\prime }Y^{\prime }Z^{\prime }W^{\prime }}$.
The totally symmetric spinors $\Psi _{ABCD}$ and its tilde counterpart $%
\tilde \Psi _{X^{\prime}Y^{\prime }Z^{\prime }W^{\prime }}$ are the Weyl
spinors. By means of the relations (\ref{simsp1})-(\ref{conrel}) we find
that the Ricci spinor is of the form 
\begin{equation}
\label{trspin1}R_{AX^{\prime }BY^{\prime }}=-2\Phi _{ABX^{\prime }Y^{\prime
}}+6\Lambda \,\epsilon _{AB}\,\epsilon _{X^{\prime }Y^{\prime }} 
\end{equation}
where we note again that there is no independent tilde version of the
trace-free Ricci spinor $\Phi _{ABX^{\prime }Y^{\prime }}$. Thus the
existence of two independent spin-frames naturally leads to the definition
of two independent Weyl spinors while the Ricci spinor remains uneffected.
The curvature complex dyad scalars are related to the curvature spinors as follows
\begin{equation}
\label{npnot}
\begin{array}{lll}
\Phi _{00}:=\Phi _{000^{\prime }0^{\prime }}\qquad & \Phi _{01}:=\Phi
_{000^{\prime }1^{\prime }}\qquad & \Phi _{02}:=\Phi _{001^{\prime
}1^{\prime }} \\ 
\Phi _{10}:=\Phi _{010^{\prime }0^{\prime }} & \Phi _{11}:=\Phi
_{010^{\prime }1^{\prime }} & \Phi _{12}:=\Phi _{011^{\prime }1^{\prime }}
\\ 
\Phi _{20}:=\Phi _{110^{\prime }0^{\prime }} & \Phi _{21}:=\Phi
_{110^{\prime }1^{\prime }} & \Phi _{22}:=\Phi _{111^{\prime }1^{\prime }} 
\end{array}
\end{equation}
\vspace{2mm} 
\begin{equation}
\begin{array}{lll}
\Psi _0:=\Psi _{0000}\qquad \quad & \Psi _1:=\Psi _{0001}\qquad \quad & \Psi
_2:=\Psi _{0011} \\ 
\Psi _3:=\Psi _{0111} & \Psi _4:=\Psi _{1111} &  \\ 
[3mm] \tilde \Psi _0:=\tilde \Psi _{0^{\prime }0^{\prime }0^{\prime
}0^{\prime }} & \tilde \Psi _1:=\tilde \Psi _{0^{\prime }0^{\prime
}0^{\prime }1^{\prime }} & \tilde \Psi _2:=\tilde \Psi _{0^{\prime
}0^{\prime }1^{\prime }1^{\prime }} \\ 
\tilde \Psi _3:=\tilde \Psi _{0^{\prime }1^{\prime }1^{\prime }1^{\prime }}
& \tilde \Psi _4:=\tilde \Psi _{1^{\prime }1^{\prime }1^{\prime }1^{\prime
}} &  
\end{array}
\end{equation}
and their relationship to the complex dyad components of the Riemann tensor are
given by eqs.(\ref{weylsc1})-(\ref{scalarcurv}) which also shows that 
\begin{equation}
\Psi _0 = \overline \Psi _4 \,,\qquad \Psi _1 = - \overline \Psi _3 \,,
\qquad \tilde \Psi _0 = \overline{\tilde \Psi} _4 \,, \qquad \tilde \Psi _1
= - \overline {\tilde \Psi} _3 \;.
\label{weylrel}
\end{equation}
Under the action of the tilde operation the Weyl and the trace-free Ricci
spinors undergo the transformation 
\begin{eqnarray}
\Psi _0 \;\tilde \leftrightarrow \;\tilde \Psi _4 \,, \qquad  &
\Psi _1\; \tilde \leftrightarrow \;\tilde \Psi _3 \,, \qquad   &
\Psi _2 \; \tilde \leftrightarrow \; \tilde \Psi _2 \, \label{weylstilde} \\
\Phi _{00} \; \tilde \leftrightarrow \;\Phi _{22}\,,  \qquad     &
\Phi _{01} \; \tilde \leftrightarrow \; \Phi _{12} \,,  \qquad     &
\Phi _{10} \; \tilde \leftrightarrow \; \Phi _{21}    \nonumber
\end{eqnarray}
with $\Phi _{11} \,, \Phi _{02}\,, \Phi _{20} $ uneffected.

\subsection{Self-dual curvature}

We shall now show that the two independent Weyl spinors $\Psi _{ABCD}$ and $%
\tilde \Psi _{X^{\prime }Y^{\prime }Z^{\prime }W^{\prime }}$ determine the
self-dual and anti-self-dual parts of curvature respectively. Beginning with
the spinor equivalent of the totally anti-symmetric Levi-Civita tensor
density (\ref{lcspin'}) we find that the spinor analog of the dual $%
^{*}R_{\alpha \beta \gamma \delta }=\frac 12\,{\epsilon _{\alpha \beta }}%
^{\mu \nu }\,R_{\mu \nu \gamma \delta}$ is given by
\begin{equation}
\label{dualcursp}^{*}R_{AX^{\prime }BY^{\prime }CZ^{\prime }DW^{\prime }}=%
\textstyle{\frac{1}{2}}\,\epsilon _{AX^{\prime }BY^{\prime }}^{\qquad \quad
ES^{\prime }FT^{\prime }}\ R_{ES^{\prime }FT^{\prime }CZ^{\prime }DW^{\prime
}}=R_{AY^{\prime }BX^{\prime }CZ^{\prime }DW^{\prime }} 
\end{equation}
which simply flips the primed indices in the first pair. Using this property
for the self-dual part of the curvature spinor (\ref{curdecomsp}) we find 
\begin{equation}
\label{sdcursp}
\begin{array}{rcl}
^{+}R_{AX^{\prime }BY^{\prime }CZ^{\prime }DW^{\prime }} & = & \frac 12 \,
\left( R_{AX^{\prime }BY^{\prime }CZ^{\prime }DW^{\prime }} +
\;^{*}R_{AX^{\prime }BY^{\prime }CZ^{\prime }DW^{\prime }}\right) \\ 
[2mm] & = & \,^{+}R_{[AB](X^{\prime }Y^{\prime })\,CZ^{\prime }DW^{\prime }}
\\ 
[2mm] & = & \tilde \Psi _{X^{\prime }Y^{\prime }Z^{\prime }W^{\prime
}}\,\epsilon _{AB}\,\epsilon _{CD}+\,\Phi _{X^{\prime }Y^{\prime
}CD}\,\epsilon _{AB}\,\epsilon _{Z^{\prime }W^{\prime }} \\ 
[2mm] &  & \ +\ \Lambda \,\epsilon _{AB}\,\epsilon _{CD}(\epsilon
_{X^{\prime }Z^{\prime }}\,\epsilon _{Y^{\prime }W^{\prime }}+\epsilon
_{X^{\prime }W^{\prime }}\,\epsilon _{Y^{\prime }Z^{\prime }}) 
\end{array}
\end{equation}
while the anti-self-dual curvature spinor is given by 
\begin{equation}
\label{asdcursp}
\begin{array}{rcl}
^{-}R_{AX^{\prime }BY^{\prime }CZ^{\prime }DW^{\prime }} & = & \frac 12\,
\left( R_{AX^{\prime }BY^{\prime }CZ^{\prime }DW^{\prime
}}-\;^{*}R_{AX^{\prime }BY^{\prime }CZ^{\prime }DW^{\prime }}\right) \\ 
[2mm] & = & \,^{-}R_{(AB)[X^{\prime }Y^{\prime }]\,CZ^{\prime }DW^{\prime }}
\\ 
[2mm] & = & \Psi _{ABCD}\,\epsilon _{X^{\prime }Y^{\prime }}\,\epsilon
_{Z^{\prime }W^{\prime }}\,+\,\Phi _{ABY^{\prime }Z^{\prime }}\,\epsilon
_{X^{\prime }Y^{\prime }}\,\epsilon _{CD} \\ 
[2mm] &  & +\Lambda \,\epsilon _{X^{\prime }Y^{\prime }}\,\epsilon
_{Z^{\prime }W^{\prime }}(\epsilon _{AC}\,\epsilon _{BD}+\epsilon
_{AD}\,\epsilon _{BC}). 
\end{array}
\end{equation}
From the above decompositions it is evident that the necessary and
sufficient conditions for the spinor of curvature to be self-dual, or
anti-self-dual are given by either 
\begin{equation}
\label{sdspincond}\Psi _{ABCD}\equiv 0,\;\;\;\;\;\Phi _{ABZ^{\prime
}W^{\prime }}\equiv 0,\;\;\;\;\Lambda \equiv 0, 
\end{equation}
or 
\begin{equation}
\label{asdspincond}\tilde \Psi _{X^{\prime }Y^{\prime }Z^{\prime }W^{\prime
}}\equiv 0,\quad \;\;\Phi _{X^{\prime }Y^{\prime }CD}\equiv 0\;\;\;\;\Lambda
\equiv 0 
\end{equation}
which illustrates the well known fact that (anti)-self-duality of the
Riemann curvature tensor implies Ricci-flatness and is a much stronger
requirement than Ricci-flatness itself.

\section{Curvature 2-forms}

For gravitational instantons curvature is an ${\cal SO}_4$-valued $2$-form.
The spinor equivalent of the Riemann curvature $2$-form splits naturally
into its self-dual and anti-self-dual parts 
\begin{equation}
\label{cur2form}
\begin{array}{rcl}
\theta _{\alpha \beta} \;\;\leftrightarrow \;\; \theta _{AX^{\prime
}BY^{\prime }} & = & \textstyle{\frac{1}{2}}\, R_{AX^{\prime }BY^{\prime
}CZ^{\prime }DW^{\prime}} \, \sigma ^{CZ^{\prime }}\wedge \sigma
^{DW^{\prime }} \\ [2mm] & = & \Theta _{AB} \,\epsilon _{X^{\prime
}Y^{\prime }} + \tilde \Theta _{X^{\prime }Y^{\prime}} \,\epsilon _{AB} 
\end{array}
\end{equation}
that is, we have the two independent symmetric matrices of curvature 2-forms 
\begin{equation}
\label{asdcur2}
\begin{array}{rcl}
\Theta _A^{\;\;B} & = & \frac 14\, R_{AX^{\prime }\;CZ^{\prime}DW^{\prime }}
^{\;\;\;BX^{\prime }} \sigma ^{CZ^{\prime }}\wedge \sigma ^{DW^{\prime }} \\ 
[3mm] \tilde \Theta _{X^{\prime }}^{\;\;Y^{\prime }} & = & \frac 14 \,
R_{AX^{\prime } \;CZ^{\prime} DW^{\prime }}^{\;\;\;AY^{\prime }}\, \sigma
^{CZ^{\prime }}\wedge \sigma ^{DW^{\prime }}\,. 
\end{array}
\end{equation}
Using the equation (\ref{curdecomsp}) together with eqs.(\ref{hodgelab}) we
find 
\begin{eqnarray}
\label{hodgerab}
\Theta_{AB} +  ^{*}\Theta _{AB} & =&
2\, \Phi_{ABX^{\prime }Y^{\prime }} \,L^{X^{\prime }Y^{\prime }}
\nonumber \\
\Theta_{AB} -  ^{*}\Theta _{AB} & =&
2\,\left[\Psi _{ABCD}+\Lambda \,(\epsilon _{AC}\,\epsilon_{BD}+
\epsilon _{AD}\,\epsilon _{BC}) \right]\,L^{CD}
\end{eqnarray}
so that the components of the spinor valued curvature 2-form $%
\Theta_A^{\;\;B} $ in terms of the various complex dyad scalars are given by
\begin{equation}
\label{sdsp2com}
\begin{array}{lll}
{\Theta _0}^0 & = & -(\,\Psi _2-\Lambda +\Phi _{11}\,)\,l\wedge \bar
l-(\,\Psi _2-\Lambda -\Phi _{11}\,)\,m\wedge \bar m \\ 
[2mm] &  & -\bar \Psi _1\,l\wedge m+\Psi _1\,\bar l\wedge \bar m-\Phi
_{10}\,\bar l\wedge m-\Phi _{12}\,l\wedge \bar m, \\ 
[3mm] {\Theta _0}^1 & = & (\,\Psi _1+\Phi _{01}\,)\,l\wedge \bar l+(\,\Psi
_1-\Phi _{01}\,)\,m\wedge \bar m+\Phi _{02}\,l\wedge \bar m \\ 
[2mm] &  & -(\,\Psi _2+2\Lambda \,)\,l\wedge m-\Psi _0\,\bar l\wedge \bar
m+\Phi _{00}\,\bar l\wedge m, \\ 
[3mm] {\Theta _1}^0 & = & (\,\bar \Psi _1-\Phi _{21}\,)\,l\wedge \bar
l+(\,\bar \Psi _1+\Phi _{21}\,)\,m\wedge \bar m-\Phi _{22}\,l\wedge \bar m
\\ 
[2mm] &  & +\bar \Psi _0\,l\wedge m+(\,\Psi _2+2\Lambda \,)\,\bar l\wedge
\bar m-\Phi _{20}\,\bar l\wedge m=-\bar \Theta _0^{\;1} 
\end{array}
\end{equation}
and similarly 
\begin{equation}
\label{asdsp2com}
\begin{array}{lll}
{\tilde \Theta _{0^{\prime }}}^{\;\;0^{\prime }} & = & -(\,\tilde \Psi _2 -
\Lambda + \Phi _{11}\,)\,l\wedge \bar l + (\,\tilde \Psi _2 - \Lambda -\Phi
_{11}\,)\,m\wedge \bar m \\ 
[2mm] &  & -\tilde \Psi _1\,\bar l\wedge m+ 
\overline{\tilde \Psi} _1\, l\wedge \bar m + \Phi _{01}\,\bar l\wedge \bar m
+ \Phi _{21}\,l\wedge m, \\ [3mm] {\tilde \Theta _{0^{\prime }}}%
^{\;\;1^{\prime }} & = & (\,\tilde \Psi _1+\Phi _{10}\,)\,l\wedge \bar l -
(\,\tilde \Psi _1 -\Phi_{10}\,)\,m\wedge \bar m -\Phi _{20}\,l\wedge m \\ 
[2mm] &  & + (\,\tilde \Psi _2+2\Lambda \,)\,l\wedge \bar m + \tilde \Psi
_0\,\bar l\wedge m - \Phi_{00}\,\bar l\wedge \bar m, \\ 
[3mm] {\tilde \Theta _{1^{\prime }}}^{\;\;0^{\prime }} & = & (\, 
\overline{\tilde \Psi} _1 - \Phi _{12}\,)\,l\wedge \bar l -(\, \overline{%
\tilde \Psi} _1 + \Phi _{12}\,)\,m\wedge \bar m + \Phi _{22}\,l\wedge m \\ 
[2mm] &  & - \overline{\tilde \Psi} _0\,l\wedge \bar m - (\,\tilde \Psi
_2+2\Lambda \,)\,\bar l\wedge m + \Phi _{02}\,\bar l\wedge \bar m = -
\overline{\tilde \Theta} _{0^{\prime }}^{\;\;1^{\prime }} 
\end{array}
\end{equation}
where we note that under the tilde operation eqs.(\ref{sdsp2com}) and (\ref
{asdsp2com}) go into each other according to the rules (\ref{indexsymmetry}%
), (\ref{weylstilde}).

\section{Ricci identities}

The curvature $2$-forms (\ref{sdsp2com}) are related to the connection $1$%
-forms (\ref{potform}) through 
\begin{equation}
\label{defcurform}{\Theta _a}^b=d\,{\Gamma _a}^b-{\Gamma _a}^c\wedge {%
\Gamma_c}^b 
\end{equation}
and of course the same relation holds for the tilde quantities as well. The
full set of Ricci identities is the explicit form of these definitions of
curvature in terms of the spin coefficients and curvature complex dyad scalars. We
find that eqs.(\ref{defcurform}) result in the Ricci identities
\begin{equation}
\label{ricciid1}
\begin{array}{l}
D\sigma -\delta \kappa +\kappa \,(\tau +\beta +3\bar \alpha -\pi )-\sigma
\,(\rho +3\epsilon +\bar \gamma -\bar \lambda )=\Psi _0 \\ 
[2mm] D\bar \alpha -\delta \epsilon -\kappa \,(\bar \epsilon +\bar \rho
)-\bar \alpha \,(\bar \gamma -\bar \lambda )-\sigma \,(\alpha +\bar \tau
)-\epsilon \,(\pi -\beta )=\Psi _1 \\ 
[2mm] \delta \alpha +\bar \delta \bar \alpha +\epsilon \,(\lambda +\bar \rho
)+\bar \epsilon \,(\bar \lambda +\rho )+\alpha \,(\bar \alpha -\beta )+\bar
\alpha \,(\alpha -\bar \beta )+\rho \bar \rho -\sigma \bar \sigma \\ 
[2mm] \;\; =-\Psi _2+\Phi _{11}+\Lambda \\ 
[2mm] D\bar \epsilon +\bar D\epsilon -\epsilon \,(\gamma -\bar \epsilon
)-\bar \epsilon \,(\bar \gamma -\epsilon )+\alpha \,(\pi +\tau )+\bar \alpha
\,(\bar \pi +\bar \tau )+\tau \bar \tau -\kappa \bar \kappa \\ 
[2mm] \;\; =-\Psi _2-\Phi _{11}+\Lambda \\ 
[2mm] D\bar \rho +\delta \bar \tau -\bar \tau \,(\beta -\pi -\bar \alpha
)-\bar \rho \,(\bar \gamma -\bar \lambda -\epsilon )-\kappa \bar \kappa
-\sigma \bar \sigma =-\Psi _2-2\Lambda \\ 
[2mm] D\tau -\bar D\kappa -\tau \,(\bar \gamma +\epsilon )+\kappa \,(\gamma
-3\bar \epsilon )-\rho \,(\pi +\tau )-\sigma \,(\bar \pi +\bar \tau ) =\Psi
_1+\Phi _{01} \\ 
[2mm] \delta \rho +\bar \delta \sigma +\sigma \,(3\alpha -\bar \beta
)+\kappa \,(\lambda +\bar \rho )-\rho \,(\bar \alpha +\beta )-\tau \,(\rho
+\bar \lambda ) =\Phi _{01}-\Psi _1 \\ 
[2mm] \bar D\bar \alpha +\delta \bar \epsilon -\bar \epsilon \,(\tau -\beta
-\bar \alpha )+\bar \alpha \,(\bar \epsilon -\bar \rho +\gamma )+\alpha \mu
-\epsilon \,\nu -\tau \bar \rho -\sigma \bar \kappa =-\Phi _{12} \\ 
[2mm] D\rho +\bar \delta \kappa +\rho \,(\bar \gamma -\rho -\epsilon
)-\kappa \,(\bar \tau -\bar \beta -3\alpha )+\tau \bar \nu +\sigma \,\bar
\mu =\Phi_{00} \\ 
[2mm] \delta \tau -\bar D\sigma -\tau \,(\tau -\beta +\bar \alpha )-\sigma
\,(3\bar \epsilon -\bar \rho +\gamma )-\rho \mu +\kappa \,\nu =\Phi _{02} 
\end{array}
\end{equation}
and the remaining set which follows from the relation
\begin{equation}
\label{defcurform'}{\tilde \Theta _{x^{\prime}}}^{\;\;y^{\prime}}=
d\,{\tilde \Gamma_{x^{\prime}}}^{\;\;y^{\prime}}-
{\tilde \Gamma _{x^{\prime}}}^{\;\;z^{\prime}}
\wedge {\tilde \Gamma_{z^{\prime}}}^{\;\;y^{\prime}}
\end{equation}
is given by 
\begin{equation}
\label{ricciid2}
\begin{array}{l}
\bar \delta \bar \nu -D\bar \mu -\bar \nu \,(\bar \tau -\bar \pi -3\bar
\beta -\alpha )-\bar \mu \,(\epsilon -\rho +3\bar \gamma +\bar \lambda
)=\tilde {\Psi _0} \\ 
[2mm] D\bar \beta -\bar \delta \bar \gamma +\bar \beta \,(\epsilon -\rho
)+\bar \gamma \,(\bar \tau -\alpha )+\bar \nu \,(\lambda +\gamma )+\bar \mu
\,(\pi +\beta )=\tilde {\Psi _1} \\ 
[2mm] D\gamma +\bar D\bar \gamma -\gamma \,(\bar \gamma -\epsilon )-\bar
\gamma \,(\gamma -\bar \epsilon )-\bar \beta \,(\pi +\tau )-\beta \,(\bar
\pi +\bar \tau )-\pi \bar \pi +\,\nu \bar \nu \\ 
[2mm] \;\; =\tilde {\Psi _2}+\Phi _{11}-\Lambda \\ 
[2mm] \delta \bar \beta +\bar \delta \beta +\beta \,(\alpha -\bar \beta
)+\bar \beta \,(\bar \alpha -\beta )-\gamma \,(\bar \lambda +\rho )-\bar
\gamma \,(\lambda +\bar \rho )-\lambda \bar \lambda+\,\mu \bar \mu \\ 
[2mm] \;\; =\tilde {\Psi _2}-\Phi _{11}-\Lambda \\ 
[2mm] \delta \bar \pi +\bar D\bar \lambda -\bar \pi \,(\tau +\beta -\bar
\alpha )+\bar \lambda \,(\bar \epsilon -\bar \rho -\gamma )+\mu \bar \mu
+\,\nu \bar \nu =\tilde {\Psi _2}+2\Lambda \\ 
[2mm] D\bar \pi -\bar D\bar \nu +\bar \pi \,(\bar \gamma +\epsilon )+\bar
\nu \,(3\gamma -\bar \epsilon )+\bar \mu \,(\pi +\tau )+\bar \lambda \,(\bar
\pi +\bar \tau ) =\tilde {\Psi _1}+\Phi _{10} \\ 
[2mm] \delta \bar \mu +\bar \delta \bar \lambda +\bar \mu \,(\bar \alpha
-3\beta )+\bar \lambda \,(\alpha +\bar \beta )+\bar \pi \,(\bar \lambda
+\rho )-\bar \nu \,(\lambda +\bar \rho ) =-\tilde {\Psi _1}+\Phi _{10} \\ 
[2mm] \delta \bar \gamma +D\beta -\bar \gamma \,(\bar \alpha +\beta -\pi
)-\beta \,(\epsilon +\bar \gamma -\bar \lambda )-\sigma \bar \beta +\pi \bar
\lambda +\mu \bar \nu +\,\gamma \kappa =\Phi _{01} \\ 
[2mm] D \bar\lambda + \delta \bar \nu - \bar \lambda \, (\epsilon - \bar
\gamma - \bar \lambda ) - \bar \nu \,( \bar \alpha + 3 \beta - \pi ) - \bar
\mu \sigma - \kappa \bar \pi = \Phi _{00} \\ 
[2mm] D \mu - \delta \pi - \mu \,(\epsilon + 3 \bar \gamma - \bar \lambda )
+ \pi \,( \bar \alpha - \beta - \pi ) +\kappa \nu - \lambda \sigma = \Phi
_{02} . 
\end{array}
\end{equation}

We note that the last two equations in each set are equivalent, thus eqs.(%
\ref{ricciid1}) and (\ref{ricciid2}) together with their complex conjugates
consist of a full set of Ricci identities. Furthermore these identities are
related to each other by the tilde operation according to eqs.(\ref
{tildesymmetrycoef}), (\ref{weylstilde}).

\section{Bianchi identities}

Bianchi had noted \cite{bianchiid} two sets of identities satisfied by the
Riemann tensor. The cyclic identity 
\begin{equation}
\label{bianchifirst}e^b\wedge {\theta _b}^a=0 
\end{equation}
is the first one of Bianchi's identites. At this point it is worth
noting that, quite generally, the left hand side of Einstein field
equations are given by the $3$-form
\begin{equation}
\label{einstein}e^b\wedge \,^{*}{\theta _b}^a=0 
\end{equation}
so that when the curvature $2$-form is self-dual, it is the first Bianchi
identity that assures Ricci-flatness. This is unlike the case of Maxwell
field where one half of Maxwell equations which is the analog of second
Bianchi identities implies the satisfaction of source-free Maxwell equations
for a self-dual Maxwell $2$-form.

The second Bianchi identities split into two independent sets which are
given by 
\begin{equation}
\label{bianchi}{d\Theta _a}^b+{\Theta _a}^c\wedge {\Gamma _c}^b-{\Gamma _a}%
^c\wedge {\Theta _c}^b=0 
\end{equation}
together with its tilde version corresponding to the two spin frames.
Inserting the expressions (\ref{sdsp2com}) and (\ref{potform}) into eq.(\ref
{bianchi}) we find the first set of second Bianchi identities 
\begin{equation}
\label{bianchi1}
\begin{array}{rl}
(\bar D+2\bar \epsilon -2\bar \rho )\,\Psi _1-(\delta -3\tau )\,\Psi_2
+2\sigma \bar \Psi _1-\bar \kappa \,\Psi _0 +(D-2\bar \gamma +\bar \lambda
)\,\Phi_{12} &  \\ 
-(\delta +2\pi )\,\Phi _{11}+\delta \Lambda +\mu \Phi _{10} +\,\kappa \Phi
_{22}-\sigma \Phi _{21}-\bar \tau \,\Phi _{02}-\bar \rho \,\Phi _{01} & = 0
\\ 
[2mm] (\bar D+2\bar \epsilon -2\bar \rho )\,\Psi _1-(\delta -3\tau )\,\Psi_2
+2\sigma \bar \Psi _1-\bar \kappa \,\Psi _0 -(\bar D+ 2\bar \epsilon +2
\lambda )\,\Phi_{01} &  \\ 
-(\bar \delta +\bar \pi+2 \alpha -2 \bar \beta )\,\Phi _{02} -2 \delta
\Lambda -2 \tau \Phi _{11} +\nu \Phi _{00} + 2 \rho \Phi _{12} & = 0 \\ 
[2mm] (\delta -4\tau -2\bar \alpha )\,\Psi _1-(\bar D+4\bar \epsilon -\bar
\rho )\,\Psi _0 +3\sigma \Psi _2+\,(\delta +2\pi -2\bar \alpha )\,\Phi _{01}
&  \\ 
-(D-2\epsilon -2\bar \gamma +\bar \lambda )\,\Phi _{02} -\mu \Phi
_{00}-2\kappa \Phi _{12}+2\sigma \Phi _{11} & = 0 \\ 
[2mm] (D-4\rho -2\epsilon )\,\Psi _1+(\bar \delta +4\alpha -\bar \tau
)\,\Psi_0 +3\kappa \Psi _2 + (\delta +\pi -2\bar \alpha -2\beta )\,\Phi
_{00} &  \\ 
-(D-2\epsilon +2\bar \lambda )\,\Phi _{01} -2\kappa \Phi _{11}+2\sigma \Phi
_{10}-\bar \nu \,\Phi _{02} & = 0 \\ 
[2mm] (\bar \delta +2\alpha -2\bar \tau )\,\Psi _1+(D-3\rho )\,\Psi _2 -2
\kappa \bar \Psi _1-\bar \sigma \,\Psi _0 +(\delta +\pi -2\beta )\,\Phi
_{10} &  \\ 
-(D+2\bar \lambda )\,\Phi_{11} - D \Lambda -\kappa \Phi _{21} +\,\sigma \Phi
_{20}-\bar \nu \,\Phi _{12} +\bar \tau \,\Phi _{01}+\bar \rho \Phi _{00} & =
0 \\ 
[2mm] (\bar \delta +2\alpha -2\bar \tau )\,\Psi _1+(D-3\rho )\,\Psi _2 -2
\kappa \bar \Psi _1-\bar \sigma \,\Psi _0 + (\bar \delta +2 \bar \pi + 2
\alpha) \,\Phi_{01} &  \\ 
+(\bar D + \lambda + 2 \bar \epsilon -2 \gamma )\,\Phi _{00} + 2 D \Lambda -
2 \rho \Phi _{11} + \bar \mu \,\Phi _{02} +2 \tau \Phi _{10} & = 0 \\ 
[2mm] &  
\end{array}
\end{equation}
The remaining set of Bianchi identities is the tilde version
of eqs.(\ref{bianchi}). In terms of the spin coefficients and complex dyad scalars
they are given by
\begin{equation}
\begin{array}{rl}
\label{bianchi2} (\bar D+2\lambda -2\gamma )\,\tilde \Psi _1+(\bar \delta
+3\bar \pi )\,\tilde \Psi _2 - 2\bar \mu \,\overline{\tilde \Psi} _1-\nu \,\tilde \Psi _0
+(D+2\epsilon -\rho )\,\Phi_{21} &  \\ 
+\,(\bar \delta -2\bar \tau )\,\Phi _{11} -\bar \delta \Lambda -\pi \Phi
_{20} +\,\lambda \Phi _{10}-\bar \sigma \Phi _{01}+\bar \mu \Phi _{12}+\bar
\nu \Phi _{22} & = 0 \\ 
[2mm] (\bar D+2\lambda -2\gamma )\,\tilde \Psi _1+(\bar \delta +3\bar \pi
)\,\tilde \Psi _2-2\bar \mu \,\overline{\tilde \Psi} _1-\nu \,\tilde \Psi _0 -(\bar D-
2 \gamma -2 \bar \rho )\,\Phi_{10} &  \\ 
+\,(\delta + 2 \bar \alpha -2 \beta - \tau )\,\Phi _{20} + 2 \bar \delta
\Lambda + \bar \kappa \Phi _{00} - 2 \bar \pi \Phi _{11}- 2 \bar \lambda
\Phi _{21} & = 0 \\ 
[2mm] (\bar \delta +4\bar \pi +2\bar \beta )\,\tilde \Psi _1+(\bar D-4\gamma
+\lambda )\,\tilde {\Psi _0}+3\bar \mu \,\tilde \Psi _2 +\,(\bar \delta
+2\bar \beta -2\bar \tau )\,\Phi _{10} &  \\ 
+(D+2\epsilon +2\bar \gamma -\rho )\,\Phi _{20} -\bar \sigma \,\Phi
_{00}+2\bar \nu \,\Phi _{21} +2\bar \mu \,\Phi _{11} & = 0 \\ 
[2mm] (D+4\bar \lambda +2\bar \gamma )\,\tilde \Psi _1-(\delta -4\beta +\pi
)\,\tilde \Psi _0+3\bar \nu \,\tilde \Psi _2 -(D-2\rho +2\bar \gamma )\,\Phi
_{10} &  \\ 
-\,(\bar \delta +2\bar \beta +2\alpha -\bar \tau )\,\Phi _{00} -\kappa \Phi
_{20}-2\bar \nu \,\Phi _{11}-2\bar \mu \,\Phi _{01} & = 0 \\ 
[2mm] (\delta +2\pi -2\beta )\,\tilde \Psi _1-(D+3\bar \lambda )\, \tilde
\Psi_2+2\bar \nu \,\overline{\tilde \Psi}_1-\mu \,\tilde \Psi _0
+\,(\bar \delta -\bar
\tau +2\alpha )\,\Phi _{01} &  \\ 
+(D-2\rho )\,\Phi_{11} +D\Lambda +\kappa \Phi _{21} +\,\lambda \Phi
_{00}-\pi \Phi _{10}+\bar \nu \,\Phi _{12}+\bar \mu \,\Phi _{02} & = 0 \\ 
[2mm] (\delta +2\pi -2\beta )\,\tilde \Psi _1-(D+3\bar \lambda )\, \tilde
\Psi_2+2\bar \nu \,\overline{\tilde \Psi}_1-\mu \,\tilde \Psi _0 +
 \,(\delta - 2 \beta
- 2 \tau )\,\Phi _{10} &  \\ 
- (\bar D + 2 \bar \epsilon - 2 \gamma - \bar \rho )\,\Phi_{00} -2 D\Lambda
+\sigma \Phi _{20} - 2 \bar \pi \Phi _{01} - 2 \bar \lambda \,\Phi_{11} & =
0
\end{array}
\end{equation}
Looking at the first and last pair of equations in both sets of eqs.(\ref
{bianchi1}) and (\ref{bianchi2}) we note that in each pair one equation is
equivalent to the other, so that we have only the eight identites which
together with their complex conjugates provide us with the full set of the
Bianchi identities. Furthermore, from the first and last pair of equations
in (\ref{bianchi1}) and (\ref{bianchi2}) we obtain the contracted Bianchi
identities involving only the Ricci scalars
\begin{eqnarray}
\label{invricci1}
(D-2\bar \gamma +\bar \lambda)\,\Phi_{12} -(\delta +2\pi )\,\Phi _{11}+
\delta \Lambda +\mu \Phi _{10} \nonumber \\
 +\,\kappa \Phi_{22}
-\sigma \Phi _{21}-\bar \tau \,\Phi _{02}-\bar \rho \,\Phi _{01} & &
 \nonumber \\
= -(\bar D+ 2\bar \epsilon +2\lambda )\,\Phi_{01}
 -(\bar \delta +\bar \pi+2 \alpha -2 \bar \beta )\,\Phi _{02}
 \nonumber \\
 -2 \delta\Lambda -2 \tau \Phi _{11} +\nu \Phi _{00} + 2 \rho \Phi _{12}
\\ [2mm]
(\delta +\pi -2\beta )\,\Phi_{10}-(D+2\bar \lambda )\,\Phi_{11} - D \Lambda
-\kappa \Phi _{21}
\nonumber \\
+\,\sigma \Phi_{20}-\bar \nu \,\Phi _{12}
+\bar \tau \,\Phi _{01}+\bar \rho \Phi _{00}
& &
\nonumber \\
= (\bar \delta +2 \bar \pi + 2 \alpha) \,\Phi_{01}
+(\bar D + \lambda + 2 \bar \epsilon -2 \gamma )\,\Phi _{00}
\nonumber \\
+ 2 D \Lambda -
2 \rho \Phi _{11} + \bar \mu \,\Phi _{02} +2 \tau \Phi _{10} \nonumber
\end{eqnarray}
We note that the tilde operation swaps both sets of the above Bianchi
identities according to eqs.(\ref{tildesymmetrycoef}), (\ref{weylstilde}).

\section{Maxwell field}

The discussion of the (anti)-self-dual properties of the electromagnetic
field in $4$-dimensional Riemannian manifold with positive definite
signature follows closely the general scheme we have presented for the Ricci
and second Bianchi identities. In particular, the source-free Maxwell
equations also fall into two independent sets. To show this we start by the
decomposition of the Maxwell $2$-form 
\begin{equation}
\label{max2form}F=\frac 12\,F_{\mu \nu }dx^\mu \wedge dx^\nu = \frac
12\,F_{AX^{\prime}BY^{\prime}} \,\sigma^{AX^{\prime}} \wedge
\sigma^{BY^{\prime}} 
\end{equation}
into its self-dual and anti-self-dual parts using, once again, the relation 
\begin{equation}
\label{asdmaxexp}F_{AX^{\prime}BY^{\prime}} =
\varphi_{AB}\,\epsilon_{X^{\prime }Y^{\prime }} + \tilde \varphi_{X^{\prime
}Y^{\prime }}\,\epsilon_{AB} 
\end{equation}
where $\varphi_{AB}$ and $\tilde \varphi_{X^{\prime}Y^{\prime}}$ are
symmetric spinors, together with (\ref{bfexp}). Explicitly we find that the
(anti)-self-dual Maxwell 2-forms are given by 
\begin{eqnarray}
\label{asdmaxexpan}
{\cal F} &  = & \frac 12\,(F\,- \,^{*}F) =  \varphi_{AB}\,L^{AB}
\nonumber \\
&=&  - \varphi _1\,\left( l\wedge \bar l+m\wedge \bar m\right)
 +\varphi _0\,\bar l\wedge \bar m+\bar \varphi _0\,l\wedge m  \\ [2mm]
\widetilde {\cal F} & = & \frac 12\,(F\,+ \,^{*}F) =
\tilde \varphi_{X'Y'}\,\tilde L^{X'Y'} \nonumber \\
& = & - \tilde \varphi _1\,\left( l\wedge \bar l - m\wedge \bar m\right)
 - \tilde \varphi _0\,\bar l\wedge m  - \overline{\tilde \varphi}_0
 \, l\wedge \bar m  \nonumber
\end{eqnarray}
where we have introduced the complex dyad scalars of the Maxwell field
\begin{equation}
\label{maxsc1}
\begin{array}{c}
\varphi _1 : = \varphi_{01} = \frac 12\,F_{\mu \nu }\, \left( l^\mu \bar
l^\nu + m^\mu \bar m^\nu \right) ,\qquad \qquad \varphi _0 : = \varphi_{00}
= F_{\mu \nu } l^\mu m^\nu \\ 
[2mm] \tilde \varphi _1 : = \tilde \varphi_{0^{\prime}1^{\prime}} = \frac
12\,F_{\mu \nu }\,\left( l^\mu \bar l^\nu - m^\mu \bar m^\nu \right) ,\qquad
\qquad \tilde \varphi _0 : = \tilde \varphi_{0^{\prime}0^{\prime}} = F_{\mu
\nu } \bar m^\mu l^\nu \\ 
[2mm] \bar \varphi _0 := \varphi _{11} ,\qquad \qquad \qquad \qquad \qquad
\qquad \overline{\tilde \varphi} _0 := \tilde \varphi
_{1^{\prime}1^{\prime}} 
\end{array}
\end{equation}
which add up to the six real components of the Maxwell field. Under the
tilde operation the complex dyad scalars manifest the correspondence
\begin{equation}
\label{tildemaxsc}\varphi _1 \; \tilde \leftrightarrow \; - \tilde \varphi
_1 , \qquad \qquad \varphi _0 \; \tilde \leftrightarrow \; - \overline{%
\tilde \varphi} _0
\label{tilmax}
\end{equation}
We conclude that in the spinor formalism for the positive definite signature
the electromagnetic fields are completely described by means of two complex
( $\varphi _0$ , $\tilde \varphi _0$ ) and two purely imaginary ( $\varphi_1 
$ , $\tilde \varphi_1 $ ) complex dyad scalars.

\subsection{Source-free Maxwell equations}

In vacuum the self-dual and anti-self-dual $2$-forms (\ref{asdmaxexpan})
both satisfy the source-free Maxwell equations 
\begin{equation}
\label{maxeq1}d\,{\cal F}=0 , \qquad d\,\widetilde{{\cal F}}=0, 
\end{equation}
and from eqs.(\ref{asdmaxexpan}) and (\ref{maxeq1}) using the relations (\ref
{d}) and (\ref{dlm}) we arrive at the two independent sets of the equations 
\begin{equation}
\label{eq1}
\begin{array}{rcl}
(\delta -2\tau )\,\varphi _1-(\bar D+2\bar \epsilon -\bar \rho )\,\varphi
_0+\sigma \,\bar \varphi _0 & = & 0 \\ 
[2mm] (D-2\rho )\, \varphi _1+(\bar \delta +2\alpha -\bar \tau )\,\varphi
_0+\kappa \,\bar \varphi _0 & = & 0 
\end{array}
\end{equation}
and 
\begin{equation}
\label{eq2}
\begin{array}{rcl}
(\bar \delta +2 \bar \pi )\, \tilde \varphi _1 +( \bar D - 2 \gamma +
\lambda )\,\tilde \varphi_0 +\bar \mu \,\overline{\tilde \varphi}_0 & = & 0
\\ 
[2mm] ( D+ 2 \bar \lambda )\, \tilde \varphi _1- ( \delta -2 \beta + \pi)\,
\tilde \varphi _0 + \bar \nu \,\overline{\tilde \varphi}_0 & = & 0 
\end{array}
\end{equation}
which consist of the Newman-Penrose version of the full set of source-free
Maxwell equations in Euclidean signature. It is clear that the equations (%
\ref{eq1}) and (\ref{eq2}) go into each other under the tilde operation
according to the correspondences (\ref{tildesymmetry}),
(\ref{tildesymmetrycoef}) and  (\ref{tilmax}).

\section{Topological numbers}

   The Newman-Penrose formalism for Euclidean signature simplifies the
discussion of the topological invariants of gravitational instantons.
We shall express the Euler number and
Hirzebruch signature as integrals over the spinor equivalents of
curvature and connection on a manifold ${\cal {M}}$ whereby the
corresponding self-dual and anti-self-dual contributions become
explicitly manifest.
We begin with the Chern formulae \cite{chern}, \cite{chsim}
for the Euler number
\begin{eqnarray}
\chi & = & \frac{1}{32 \pi^2}\int_{\cal {M}}
\epsilon^{abcd}\theta_{ab} \wedge \theta_{cd}
\nonumber  \\[2mm] & &
- \frac{1}{16 \pi^2}\int_{\partial \cal {M}}
\epsilon^{abcd} \left( \,\Omega_{ab} \wedge \theta_{cd}
-\,\frac{2}{3}\,\,\Omega_{ab} \wedge \Omega_{c}^{\;\;e}\,
\wedge \Omega_{ed}\,\right)
\label{euler1}
\end{eqnarray}
and  the Hirzebruch signature
\begin{equation}
\tau  =  -\frac{1}{24 \pi^2}\int_{\cal {M}}
 \theta_{a}^{\;\;b} \wedge \theta_{b}^{\;\;a}  +
\frac{1}{24 \pi^2}\int_{\partial \cal {M}}
\Omega_{a}^{\;\;b} \wedge \theta_{b}^{\;\;a} - \eta_{s}({\partial \cal {M}})
\label{sign1}
\end{equation}
of a non-compact manifold. Here $\, \theta_{ab}\, $ is the curvature 2-form,
 $\, \Omega_{ab}\, $ is the second fundamental form of the boundary
$\,{\partial \cal {M}}\,$
\begin{equation}
  \Omega_{ab}= \omega_{ab} - (\omega_{0})_{ab}
\label{second}
\end{equation}
which is the difference between the connection 1-form $\,\omega_{ab}\,$ for
the original metric and the connection 1-form $\,(\omega_{0})_{ab} \,$
of a product (induced) metric on the boundary $\,{\partial \cal {M}}\,$.
Finally $\,\eta_{s}({\partial \cal {M}}) \,$ is the eta-invariant
\cite{hitchin}, \cite{aps}.

 Using the spinor equivalent of the totally anti-symmetric Levi-Civita
alternating symbol (\ref{lcspin'}), the connection 1-form
(\ref{dualconsp}), and the curvature 2-form (\ref{cur2form}) we obtain
the spinor expression for the Euler number
\begin{eqnarray}
\chi & = & \frac{1}{8 \pi^2}\int_{\cal {M}}
\left( \Theta_{A}^{\;\;B} \wedge \Theta_{B}^{\;\;A} -
\,\tilde{\Theta}_{A'}^{\;\;B'} \wedge \tilde{\Theta}_{B'}^{\;\;A'} \right)
\nonumber \\[2mm] & &
-\, \frac{1}{4 \pi^2}\int_{\partial \cal {M}}\left[
\, \gamma_{A}^{\;\;B} \wedge \Theta_{B}^{\;\;A} - \,
\tilde{\gamma}_{A'}^{\;\;B'} \wedge \tilde{\Theta}_{B'}^{\;\;A'}
\nonumber \right. \\[2mm] & & \left.
-\frac{2}{3}\,(\, \gamma_{A}^{\;\;B} \wedge \gamma_{B}^{\;\;E}
 \wedge  \gamma_{E}^{\;\;A} - \,
\tilde{\gamma}_{A'}^{\;\;B'} \wedge \tilde{\gamma}_{B'}^{\;\;E'}
\wedge \tilde{\gamma}_{E'}^{\;\;A'}\,) \,\right]
\label{euler2}
\end{eqnarray}
where
\begin{equation}
\gamma_{A}^{\;\;B} =  \Gamma_{A}^{\;\;B}  - (\Gamma_{0})_{A}^{\;\;B}
\label{speqsff}
\end{equation}
is the spinor equivalent of the second fundamental form. We see that
the self-dual and anti-self-dual contributions to this integral
are explicit.

    In the case of a compact manifold when the boundary terms vanish
the spinor expression of the Euler number (\ref{euler2}) can be reduced
into the form
\begin{eqnarray}
\chi & = & \frac{1}{4 \pi^2}\int_{\cal {M}}
\left[\, |\,\Psi_{0}\,|^2 + 4\,|\,\Psi_{1}\,|^2 + 3\,\Psi_{2}^2
+ |\,\tilde \Psi_{0}\,|^2
\nonumber \right. \\[3mm] & & \left.
+\, 4\, |\,\tilde \Psi_{1}\,|^2 + 3\,  \tilde \Psi_{2}^2
-2\,( \, |\,\Phi_{00}\,|^2 +  |\,\Phi_{02}\,|^2\,)
\nonumber \right. \\[3mm] & & \left.
-\, 4\,(\, |\,\Phi_{01}\,|^2 + |\,\Phi_{11}\,|^2 +  |\,\Phi_{12}\,|^2
 - 3 \Lambda^2 \,) \,\right]
 l \wedge \bar{l} \wedge m \wedge \bar{m}
\label{eulerweyl}
\end{eqnarray}
involving the Weyl and the Ricci scalars.

Similarily, the Hirzebruch signature can be expressed as integrals of the
spinor-valued curvature 2-forms and connection 1-forms. We find
the expression
\begin{eqnarray}
\tau & = & - \frac{1}{12 \pi^2}\, \left[\, \int_{\cal {M}}
\left( \Theta_{A}^{\;\;B} \wedge \Theta_{B}^{\;\;A} +
\,\tilde{\Theta}_{A'}^{\;\;B'} \wedge \tilde{\Theta}_{B'}^{\;\;A'} \right)
\nonumber \right. \\[2mm] & & \left.
\, - \int_{\partial \cal {M}}
\,(\, \gamma_{A}^{\;\;B} \wedge \Theta_{B}^{\;\;A} + \,
\tilde{\gamma}_{A'}^{\;\;B'} \wedge \tilde{\Theta}_{B'}^{\;\;A'}\, )
\,\right] \, -  \eta_{s}({\partial \cal {M}}) .
\label{sign2}
\end{eqnarray}
which, in turn, goes to the form
\begin{eqnarray}
\tau & = & - \frac{1}{6 \pi^2}\int_{\cal {M}}
\left[\, |\,\Psi_{0}\,|^2 + 4\,|\,\Psi_{1}\,|^2 + 3\,\Psi_{2}^2
 - |\,\tilde \Psi_{0}\,|^2
\nonumber \right. \\[3mm] & & \left.
-\, 4\, |\,\tilde \Psi_{1}\,|^2 - 3\,  \tilde \Psi_{2}^2\, \right]
 l \wedge \bar{l} \wedge m \wedge \bar{m} \,- \eta_{s}({\partial \cal {M}})
\label{signweyl}
\end{eqnarray}
for a compact manifold.

\section{Petrov types}

    We shall now discuss Petrov type classification of gravitational
instantons in the spinor formalism for positive definite signature
in order to point out that the basis bi-vectors used in the classification
of the Riemann tensor are obtained from the almost complex structure
vector-valued $1$-forms in hyper-K\"ahler structure by lowering the
vector index. For Petrov classification we examine the eigenspinors and
the corresponding eigenvalues of the totally symmetric Weyl spinors
(\ref{totsym}). Following to the general scheme \cite{pr} we begin
with two sets of orthonormal triads of spinors that correspond to
two independent spin frame with bases  (\ref{01}) and  (\ref{0'1'})
respectively. These triads are simply the complex structure
bi-spinors (\ref{cssp})
\begin{eqnarray}
n^{1}_{AB}& = & -\frac{i}{\sqrt{2}}\,(\, o_A o_B  -\iota_A \iota_B \,)
 \nonumber \\[2mm]
n^{2}_{AB} & = & \frac{1}{\sqrt{2}}\,(\, o_A o_B  + \iota_A \iota_B \,)
\label{triad1} \\  [2mm]
n^{3}_{AB} & = & i \sqrt{2}\, o_{(A} \iota_{B)}   \nonumber
\end{eqnarray}
where it is clear that this triad satisfies the orthonormality condition
\begin{equation}
n^{i}_{AB}\,n_{j}^{AB} = \, \delta_{j}^i \;\;\;\;\; i,j = 1,2,3
\label{triad}
\end{equation}
and similar relations hold for primed indices.
Projecting the totally symmetric Weyl spinor $\,\Psi_{ABCD}\,$
onto the triad basis (\ref{triad1}) we obtain the following
traceless  $ 3 \times 3 $ matrix
\begin{equation}
\Psi = \left(  \begin{array}{ccc}
\Psi_2 - \frac{1}{2} ( \Psi_0 + \overline \Psi_0 )  &
- \frac{i}{2} ( \Psi_0 - \overline \Psi_0 ) &  \Psi_1
+ \overline \Psi_1  \\[3mm]
- \frac{i}{2} ( \Psi_0 - \overline \Psi_0 ) &
\Psi_2 + \frac{1}{2} ( \Psi_0 + \overline \Psi_0 ) & i (\Psi_1
- \overline  \Psi_1) \\ [3mm]
 \Psi_1 + \overline  \Psi_1 & i (\Psi_1 - \overline  \Psi_1) &
 -2 \Psi_2  \end{array} \right)
\label{matrix}
\end{equation}
where we have used the relations (\ref{weylrel}). Since this matrix is real
and symmetric it can always be diagonalised in the orthonormal
basis with eigenspinors (\ref{triad1}).
Corresponding to three distinct real eigenvalues
\begin{equation}
 \lambda_{1} = \Psi_2 - \Psi_0, \qquad \qquad \lambda_{2} = \Psi_2 + \Psi_0,
\qquad \qquad   \lambda_{3} = -2 \Psi_2
\label{diagonal}
\end{equation}
obeying the condition
$$\, \lambda_{1} + \lambda_{2} + \lambda_{3} = 0  $$
we have the case of algebraically general, or Petrov type $I$,
gravitational instantons. When two roots coincide
$\, \lambda_{1} =  \lambda_{2}\, $, from eqs.(\ref{diagonal}) and
(\ref{weylrel}) we obtain
\begin{equation}
\Psi_2 = \lambda_{1} = \lambda_{2}, \qquad
\Psi_0 = \Psi_1 = \Psi_3 = \Psi_4 = 0
\label{dtype}
\end{equation}
which corresponds to the algebraically special, or Petrov type $D$,
gravitational instantons. For the anti-self-dual case a similar analysis
holds for the independent primed  $ 3 \times 3 $ matrix.  These results are
in agreement with those obtained in \cite{karl}, \cite{g2}. Thus
gravitational instantons with (anti)-self-dual curvature must be
either algebraically general of Petrov type $I$, or algebraically
special of Petrov type $D$. We have seen that the
basis bi-vectors used in the Petrov classification
of the Weyl tensor are directly related to hyper-K\"ahler structure.

\section{Conclusion}

    We have presented the NP formalism in Euclidean signature in terms
of differential forms in the complex isotropic dyad basis and its spinor
equivalent. This formalism is naturally adopted to the
discussion of the physical and mathematical properties of gravitational
instantons where the curvature is self-dual. As part of a systematic
exposition of this formalism we have presented some known results,
however, we feel that this NP formalism can be used in many problems
of current interest in both physics and mathematics. In particular,
we have presented the explicit expression for the vector-valued
$1$-forms that define three almost complex structures for a general
gravitational instanton metric which admits a self-dual curvature
$2$-form. We have found that the integrability condition for these
almost complex structures, namely the vanishing of the Nijenhuis
vector-valued $2$-form is automatically satisfied in the self-dual
gauge which is guaranteed by the condition of self-dual curvature.
This is explicit proof of hyper-K\"ahler structure.
The essential tool which makes this general result possible is
the Newman-Penrose formalism for Euclidean signature.
We have shown that the Ricci
and the Bianchi identities, as well as the Maxwell equations, naturally
fall into two independent sets which express their self-dual and
anti-self-dual content. Finally we have shown the relationship between
hyper-K\"ahler structure for gravitational instantons and
Petrov classification.

\section{Acknowledgement}

   We thank G. W. Gibbons for many interesting discussions and
correspondence.

\end{document}